\documentclass[10pt,leqno]{article}
\usepackage{latexsym}
\usepackage{amsfonts}
\usepackage{amsmath}
\def\CC {{\mathbb C}}	     
\def\RR {{\mathbb R}}	     
\def\PP {{\mathbb P}}	     
\def\ZZ {{\mathbb Z}}	     
\def\la{\,\langle\,}
\def\ra{\,\rangle\,}
\def\bp{\stackrel{{\boldsymbol{.}}}}
\def\bpp{\stackrel{{\boldsymbol{.}}{\boldsymbol{.}}}}
\def\ben{\begin{eqnarray*}}
\def\een{\end{eqnarray*}}
\def\be{\begin{eqnarray}}
\def\ee{\end{eqnarray}}
\def\ban{\begin{align*}}
\def\ean{\end{align*}}
\def\mm{\vspace{2mm}}
\def\vm{\vspace{-0.1cm}}
\def\vmm{\vspace{-0.2cm}}
\def\vmmm{\vspace{-0.3cm}}
\def\epsi{\epsilon \,}

\newtheorem{th1}{Theorem}[section]
\newtheorem{pr}[th1]{Proposition}
\newtheorem{lm}[th1]{Lemma}

\title{{\bf The Complex Geometry of Lagrange Top}}
\author{{\bf Lubomir Gavrilov}\\[3pt]
\normalsize \it Laboratoire Emile Picard, CNRS UMR 5580,
 Universit\'e Toulouse III\\
\normalsize \it 118, route de Narbonne,
 31062 Toulouse Cedex, France \\[10pt]
 {\bf Angel Zhivkov}\thanks{Supported by DFG project
 number 436 BUL 113/86/5 }\\
\normalsize \it Faculty of Mathematics and Informatics,
 Sofia	University\\
\normalsize \it 5 J. Bourchier, 1164 Sofia, Bulgaria}

\begin{document}

\maketitle

$\qquad$    MS Classification 58F07, 70E15
\begin{abstract}

We prove that the heavy symmetric top (Lagrange, 1788)
linearizes on a two--dimensional non--compact algebraic
group -- the generalized Jacobian of an elliptic curve
with two points identified. This leads to a transparent
description of its complex and real invariant level sets.
We deduce, by making use of a Baker--Akhiezer
function, simple explicit formulae for the general
solution of the Lagrange top. At last we describe the
two real structures of the Lagrange top and their relation with
the focusing and the non--focusing non--linear Schr\"odinger equation. 

\end{abstract}

\tableofcontents


\section{Introduction}

The motion under gravity of a  rigid body one of whose
points is fixed  is described by a Hamiltonian system
on the cotangent bundle ${\rm T^*SO(3)}$ of its
configuration space ${\rm SO(3)}$, coordinatized by
Euler angles and their conjugate  momenta. This system
was first obtained by Lagrange around 1788 \cite{Lagrange},
the particular case of free rigid body motion being
already known to Euler. After a first reduction,
with respect to rotations about the vertical in space,
this leads to the following two degrees of freedom
Hamiltonian system on ${\rm T^* S^2}$ obtained by
Lagrange too \cite[p.232 and p.243]{Lagrange}
\begin{equation}
\label{EP}
	\frac{dM}{dt}
      = M\times\Omega \, + \, \chi\times\Gamma , \qquad
	\frac{d\Gamma}{dt}  = \Gamma\times\Omega
\end{equation}
$$
	M = (M_1,M_2,M_3),		       \  \
       \Omega=(\Omega _1,\Omega _2,\Omega _3), \  \
       \Gamma = (\Gamma_1,\Gamma_2,\Gamma_3),  \  \
       \chi=(\chi_1,\chi_2,\chi_3) .
$$
Here $M,\Omega $ and $\Gamma$ denote respectively the angular
momentum, the angular velocity and the coordinates of the unit
vector in the direction of gravity, all expressed in
body--coordinates. The constant vector $\chi$ is the center
of mass in body--coordinates multiplied by the mass of the body
and the acceleration. We recall that $M=I\Omega $ where $I$
is the matrix of the inertia operator and we may suppose that
$I={\rm diag}(I_1,I_2,I_3)$. The system (\ref{EP}) may be viewed
as a two degrees of freedom Hamiltonian system on the manifold
${\rm se^*(3) \sim se(3)}$ --
the Lie algebra of the Euclidean group of
three space ${\rm SE(3) = SO(3)} \times \RR^3$.
Indeed, ${\rm se^*(3)}$ with
its usual Kostant--Kirillov--Poisson structure may be identified,
via (a multiple of) the Killing form, with ${\rm se(3)}$.
This induces the following Lie--Poisson bracket on
${\rm se(3)} \sim \RR^3\times \RR^3$
$$
      \{M_1,M_2\} = - M_3 \, ,...,		 \
      \{M_1,\Gamma_2\} = - \Gamma_3 \, ,...,	 \
      \{\Gamma_i,\Gamma_j\}=0
$$
with coadjoint orbits
$$
      {\cal M}_{a} = \bigl\{ \, (M,\Gamma)\in \RR^6  \, : \,
			\la \Gamma,\Gamma \ra =1,   \,
			\la \Gamma,M \ra = a  \,
		     \bigr\} \ .
$$
and on each symplectic leaf (\ref{EP}) is Hamiltonian  with
Hamiltonian function the energy of the body (see \cite{Ratiu})
$$
      E = \tfrac{1}{2}	   \la \Omega , M   \ra
			 - \la \chi,\Gamma  \ra  \; .
$$

	Further we shall be interested in the case when the
body is symmetric about an axis through the center of gravity
and the fixed point -- the so called Lagrange top
\cite[p.253]{Lagrange}. This is equivalent to the conditions
$I_1=I_2$ and $\chi=(0,0,\chi_3)$. Without loss of generality
we may also suppose that $\chi_3/I_1=1$, and if we put
$m=(I_3-I_2)/I_1$ then (\ref{EP}) takes the form
\begin{align}
       \bp{\Omega}_1
   & =	  -m \, \Omega_2 \Omega_3 - \Gamma_2  &
       \bp{\Gamma}_1  & =     \Gamma_2 \Omega_3
			    - \Gamma_3 \Omega_2
\nonumber
\\
\label{Ltop}
       \bp{\Omega}_2
   & = \ \ m \, \Omega_3 \Omega_1 + \Gamma_1  &
       \bp{\Gamma}_2  & =     \Gamma_3 \Omega_1
			    - \Gamma_1 \Omega_3
\\
\nonumber
       \bp{\Omega}_3  & = \ \	0	      &
       \bp{\Gamma}_3  & =     \Gamma_1 \Omega_2
			    - \Gamma_2 \Omega_1
\end{align}
with first integrals
\ben
     H_1 &=& \Gamma_1^2 + \Gamma_2^2 + \Gamma_3^2
\\
     H_2 &=& \Omega_1 \Gamma_1 + \Omega_2 \Gamma_2
			       + (1+m) \Omega_3 \Gamma_3
\\
       E  =  H_3
	 &=&  \tfrac{1}{2} \, \bigl( \,
	      \Omega_1^2 + \Omega_2^2 + (1+m) \Omega_3^2
			   \, \bigr)
	     - \Gamma_3 \ .
\een

\vspace{1cm}

\begin{center}
\begin{picture}(50,50)(10,50)
\setlength{\unitlength}{.4mm}

\linethickness{.6pt}
\put(4.28,30){\line(-1,0){64.28}}      
\qbezier(-60,30)(-70,0)(-80,-30)       
\put(-80,-30){\line(1,0){170}}	       
\qbezier(90,-30)(85,0)(80,30)	       
\put(80,30){\line(-1,0){50}}	       

\linethickness{1pt}
\qbezier(60,60)(30,30)(0,0)	      
\qbezier(0,0)(6,42)(12,84)	      
\qbezier(12,84)(30,60)(60,60)	      
\qbezier(60,60)(54.96,68.36)(47,74)   
\qbezier(12,84)(16.7,85)(31,82)       

\qbezier(28,76)(33,86)(38,96)	      
\qbezier(44,68)(49,78)(54,88)	      
\qbezier(54,88)(47.5,95)(38,96)       
\qbezier(54,88)(44.5,89)(38,96)       

\put(150,30){\makebox(0,0)[cc]{Fig.1: \large{Lagrange top}}}
\end{picture}
\end{center}

\vspace{3cm}

\noindent
Due to the symmetry of the body there is an additional
integral of motion
$$
     H_4 = \Omega_3
$$
which makes (\ref{Ltop}) Liouville integrable on the
symplectic leaf
$$
       {\cal M}_a = \Bigl\{ (\Omega ,\Gamma)\in \RR^6:\
	 \Gamma_1^2 + \Gamma_2^2 + \Gamma_3^2 = 1\,,\
	 \Omega_1\Gamma_1 + \Omega_2 \Gamma_2
			  + (1+m)\Omega _3\Gamma_3
		   =  a \Bigr\} \ .
$$
The Hamiltonian vector field generated by $H_4$ on ${\cal M}_a$
is given by
\begin{align}
   \bp{\Omega}_1  & =	\ \Omega_2     &
   \bp{\Gamma}_1  & =	\ \Gamma_2     &\nonumber \\
\label{Ltop1}
   \bp{\Omega}_2  & =	 -\Omega_1     &
   \bp{\Gamma}_2  & =	 -\Gamma_1     &	  \\
   \bp{\Omega}_3  & =	\   0	       &
   \bp{\Gamma}_3  & =	\   0	       &\nonumber
\end{align}
and it represents uniform rotations about the symmetry axis
through the center of gravity and the fixed point in space.

     The Lagrange top is one of the most classical examples of
integrable systems and it appears in almost all papers on this
subject. The explicit formulae for the position of the body in
space ($\Gamma_1,\Gamma_2,\Gamma_3$ in our case) were found by
Jacobi \cite[p.503--505]{Jacobi2}. In the last twenty years
most of the integrable problems of the classical mechanics
were revisited by making use of algebro--geometric techniques.
From this point of view the Lagrange top takes a somewhat
singular place -- the results available are either
incomplete, or inexact, or even wrong. Consider the complexified group of
rotations
$\, \CC^* \sim \CC/2 \pi i \ZZ$
defined by the flow of the vector field (\ref{Ltop1}). It acts freely
on the generic complex invariant level set
$$
       T_h  =  \Bigl\{			    \,
	   (\Omega ,\Gamma) \in \CC^6:	    \
       H_1 (\Omega , \Gamma) = 1\,,	    \
       H_2 (\Omega , \Gamma) = h_2\,,	    \
       H_3 (\Omega , \Gamma) = h_3\,,	    \
       H_4 (\Omega , \Gamma) = h_4	    \,
	       \Bigr\}			    
$$
and it is classically known that the quotient manifold $T_h/\CC^*$ is an
elliptic curve. The starting point of the present article is the observation
that, generically, the algebraic manifold $T_h$ {\it is not isomorphic } to a
direct product of the curve $T_h/\CC^*$ and $\CC^*$ (although
as a topological manifold it is).  Let us explain first the
algebraic structure of the invariant level set $T_h$.
If $\Lambda \subset \CC^2$ is a rank three lattice
\begin{equation}
   \label{lattice}
	  \Lambda = \ZZ\left(
			  \begin{array}{c}
			     2 \pi i \\
			     0
			  \end{array}
			\right)
	  \oplus \ZZ \left(
			  \begin{array}{c}
			     0	   \\
			     2 \pi i
			  \end{array}
		     \right)
	  \oplus \ZZ
		     \left(
			 \begin{array}{c}
			    \tau_1  \\
			    \tau_2
			 \end{array}
		     \right) ,\ \
	 {\rm Re \,}(\tau_1)<0 \;
\end{equation}
then $\CC^2/\Lambda$ is a non--compact {\it algebraic group} and
it can be considered as a (non--trivial) extension of the
elliptic curve $\,\CC/\{ 2\pi i \ZZ \oplus \tau_1 \ZZ\}\,$
by $\, \CC^* \sim \CC/2 \pi i \ZZ$
\begin{equation}
   \label{suite}
		       0    \, \rightarrow \,
	 \CC   / 2\pi i \ZZ \, \rightarrow \,
	 \CC^2 / \Lambda    \, \stackrel{\phi}{\rightarrow} \,
	 \CC   / \{ 2\pi i \ZZ \oplus \tau_1 \ZZ\}
			    \, \rightarrow 0 \, ,
   \qquad
	 \phi(z_1,z_2)=z_1 \; .
\end{equation}
We prove that, for generic $h_i$, the complex invariant level set
$T_h$ of the Lagrange top is biholomorphic to (an affine part of)
$\CC^2/\Lambda$. The algebraic group $\CC^2/\Lambda$
turns out to be the generalized Jacobian of an elliptic
curve with two points identified. This curve, say $C$, is the
spectral curve of a Lax pair for the Lagrange top, found first by
Adler and van Moerbeke \cite{Adler} and its Jacobian
${\rm Jac}(C)=\CC/\{ 2\pi i \ZZ \oplus \tau_1 \ZZ\}$ is a curve
found first by $\ldots$ Lagrange. Further we prove that the flows
(\ref{Ltop}), (\ref{Ltop1}) define translation invariant vector
fields on $\CC^2/\Lambda$ which means that our system is
algebraically completely integrable.

      Let us compare  the above to the classical Lagrange
linearization on an elliptic curve \cite{Lagrange} (see also
\cite{Adler,Ratiu,Verdier,Audin2,Audin1}). It is well known
that, due to the symmetry of the body, the system
(\ref{Ltop}) is invariant under rotations about the symmetry
axe. These rotations are given by the flow
of (\ref{Ltop1}) which commutes with the flow of the Lagrange top.
Thus we have a well defined $\CC^*$ action on the complex
invariant level set  $T_h \sim \CC^2/\Lambda$ and a well defined
(factored) flow on $T_h/\CC^*$.  Lagrange noted around
1788  that this factorization amounts to eliminate the variables
$\Omega_1, \Omega_2, \Gamma_1, \Gamma_2$ so he obtained a single
autonomous differential equation for the nutation
$\theta$, where $\Gamma_3 = \cos \theta$  \cite[p.254]{Lagrange}
(nutation is the inclination of the symmetry axis of the body to
the vertical). Finally it is seen from this equation
that $\Gamma_3(t)$ is, up to an addition and a multiplication by
a constant, the Weierstrass elliptic function $\wp(t)$. Thus
Lagrange linearized the complex flow of the Lagrange top on
an elliptic curve. This curve happens to be the Jacobian
$J(C)$ of the spectral curve $C$ of Adler and van Moerbeke
and is identified with $\CC/\{ 2\pi i \ZZ \oplus \tau_1 \ZZ\}$
in (\ref{suite}). The kernel of the map $\phi$ is just the
circle action $\CC^*\sim \CC/2\pi i \ZZ$ defined by (\ref{Ltop1}),
so the linear vector  field (\ref{Ltop1}) is projected under $\phi$
onto the zero vector field on
${\rm Jac}(C)=\CC/\{ 2\pi i \ZZ \oplus \tau_1 \ZZ\}$.

      To resume on a modern language,  Lagrange's computation shows
that {\it the generic invariant level set $T_h$ of the Lagrange top is an
extension of an elliptic  curve $C \sim {\rm Jac}(C)$ by $\CC^*$
and the flow is projected on this curve into a well defined linear flow}.
This is, however, a very vague description of $T_h \sim \CC^2/\Lambda$.
Indeed, although the fibration
\begin{equation}
	  \CC^2 / \Lambda  \,
	  \stackrel{\phi}{\rightarrow} \, {\rm Jac}(C)
       =  \CC/\{ 2\pi i \ZZ \oplus \tau_1 \ZZ\}
\end{equation}
is topologically trivial, it is not algebraically trivial, and to know
its {\it type} we need the parameter $\tau_2$ (defined in (\ref{lattice}) )
\cite{Serre}. As the general solution of (\ref{Ltop}) lives on $\CC^2/\Lambda$
then, contrary to what is often affirmed, it can not be expressed in
terms of elliptic functions and exponentials.
It is even less true  that ``the flow of the Lagrange top lives on a complex
2--dimensional cylinder with generator the line $z=0$'' as claimed in
\cite[p.232]{Ratiu}.

	The algebraic description of the Lagrange top is carried out
in section \ref{algebraic} (Theorem \ref{th.Lagrange}).
The Lax pair is  used first in section \ref{solutions} where we
construct the corresponding Baker--Akhiezer function. This implies
explicit formulae for the general solution of the Lagrange top which
complete and simplify the classical formulae due to Jacobi
\cite[p.503--505]{Jacobi2} for $\Gamma_1, \Gamma_2, \Gamma_3$
and Klein and Sommerfeld \cite[p.436]{Klein} for the angular
velocities (Theorem \ref{sigma.th}).

      In section \ref{real} we	study  reality conditions on the
(complex) solutions. Besides the usual real structure of the Lagrange
top given by complex conjugation there is a second natural
real structure induced by the eigenvalue map of the corresponding Lax
pair representation. It turns out that these two structures
coincide on  ${\rm Jac}(C)$ but are different on $\CC^2/\Lambda$
(and hence on $T_h$). The corresponding real level sets are
described in Theorem \ref{real.th}. This makes clear the
relation between the real structure of the curve $C$, its Jacobian
${\rm Jac}(C)$ and the real level set $T_h^\RR$ (a question raised
in \cite{Audin1} and \cite[p.37]{Audin2}).

      The results obtained in the present paper lead to the following
unexpected observation: the real solutions of the Lagrange top
corresponding to its two real structures provide one--gap solutions
of the	nonlinear Schr\"odinger equation (Proposition \ref{sch.pr})
$$
  \leqno{(NLS^\pm)}
  \qquad  \qquad  \qquad  \qquad  \qquad  \qquad
	u_{xx} = i u_t \pm 2 |u|^2 u	  \ .
$$

	At last, for convenience of the reader, we give in the Appendix
a brief account of some more or less well known results concerning
the linearization of the Lagrange top on an elliptic curve.


\begin{center}
{\bf Acknowledgments}
\end{center}

Part of this work was done while the second author was
visiting the University of Toulouse III in June 1994.
He is grateful for its hospitality. We also acknowledge
the interest of M.~Audin, Yu. Fedorov, V.V. Kozlov and
A. Reiman to the paper.


\section{Algebraic Structure}
\label{algebraic}

      Let $\breve{C}$ be the affine curve $\{ \mu^2=f(\lambda)\}$ where
$f$ is a degree four polynomial without double roots. We  denote by $C$
the completed and normalized curve $\breve{C}$. Thus $C$ is a compact Riemann
surface, such that $C= \breve{C} \cup \infty^+ \cup \infty^-$, where
$\infty^\pm$ are two distinct ``infinite" points on $C$.
Consider the effective divisor ${\mit
m}=\infty^++\infty^-$ on $C$ and let  $ J_m(C)$ be the generalized
Jacobian of the elliptic curve
$C$ relative to $m$. Following \cite{Serre} we shall call ${\mit m}$
a modulus. We shall
denote also $J(C;\infty^\pm)=J_m(C)$. Recall that
the usual Jacobian 
$$J(C)={\rm Div}^0(C)  \,
			 \big/ \, \sim
$$
is the additive group ${\rm Div}^0(C)$ of degree zero divisors on
$C$ modulo the equivalence relation $\sim$ . We have $D_1  \sim
D_2$ if and only if there exists a meromorphic function $f$ on $C$ such that
 $(f)=D_1-D_2$.
Similarly the generalized Jacobian
$$
       J(C;\infty^\pm) = {\rm Div}^0(\breve{C})  \,
			 \big/ \, \stackrel{m}{\sim}
$$
is the additive group ${\rm Div}^0(\breve{C})$ of degree zero divisors on
$\breve{C}$ modulo the equivalence relation $\stackrel{m}{\sim}$ .
We have $D_1 \stackrel{m}{\sim} D_2$ if and only if there exists a
meromorphic function $f$ on $C$ such that
$f(\infty^+)=f(\infty^-)=1$ and $(f)=D_1-D_2$. The generalized Jacobian
$J(C;\infty^\pm)$ is thus obtained as a $\CC^*$--extension
of the usual Jacobian $J(C)$ (isomorphic to $C$).
This means that there is an exact sequence of groups
\begin{equation}
\label{extension}
			 0   \,  \stackrel{exp}{\rightarrow}	  \,
		      \CC^*  \,  \stackrel{\upsilon}{\rightarrow} \,
	    J(C;\infty^\pm)  \,  \stackrel{\phi}{\rightarrow}	  \,
      J(C) \rightarrow 0 \; .
\end{equation}
The map $\phi$ is induced by the inclusion  $\breve{C} \subset C$ and
$\upsilon(r)\in J(C;\infty^\pm)$, $r\neq0$, is the divisor
of any meromorphic function $f$ on $C$ satisfying
$f(\infty^+)/f(\infty^-)=r$  \cite[p.55]{Fay}.

     As an analytic manifold $J(C;\infty^\pm)$ is
\begin{equation}
  \label{extension1}
       \CC^2 / \Lambda \ \sim \
	H^0 \bigl( C , \Omega^1(\infty^+ \! + \! \infty^-)
	    \bigr) ^*
	    \, \big/ \, H_1(\breve{C},\ZZ)
\end{equation}
where the lattice $\Lambda$ is generated by the three vectors
\begin{equation}
   \label{extension2}
      \Lambda_1
    = \left(
	     \begin{array}{c}
		 \int_{A_1} \frac{d\lambda}{\mu}
\\[8pt] 	 \int_{A_1}\frac{\lambda \, d\lambda}{\mu}
	     \end{array}
      \right)	, \quad
      \Lambda_2
    = \left(
	     \begin{array}{c}
		 \int_{A_2} \frac{d\lambda}{\mu}
\\[8pt] 	 \int_{A_2} \frac{\lambda \, d\lambda}{\mu}
	     \end{array}
      \right)	, \quad
      \Lambda_3
    = \left(
	     \begin{array}{c}
		 \int_{B_1} \frac{d\lambda}{\mu}
\\[8pt] 	 \int_{B_1} \frac{\lambda \, d\lambda}{\mu}
	     \end{array}
      \right)
\end{equation}
and the cycles $A_1, A_2, B_1$ form a  basis of the first homology
group $H_1(\breve{C},\ZZ)$  as on figure 2. It is seen
that the period lattice $\Lambda $ may be obtained by pinching
a non--zero homology cycle of a genus two Riemann surface
to a point $\infty^\pm$ (figure 2). This is expressed by saying
that $J(C;\infty^\pm)$ is the Jacobian of the elliptic curve $C$ with
two points $\infty^+$ and $\infty^-$ identified \cite{Fay}.
For a further use note also that
\begin{equation}
\label{projection}
      \phi : J(C;\infty^\pm) \rightarrow J(C) \,, \quad
      \phi : \CC^2/\Lambda   \rightarrow \CC / \phi(\Lambda )
\end{equation}
%
\begin{center}
\begin{picture}(60,60)(160,0)
\setlength{\unitlength}{.4mm}
\linethickness{1pt}
    \qbezier(-10,0)(-10,40)(100,40)
    \qbezier(100,40)(140,40)(150,30)
    \qbezier(150,30)(160,20)(150,10)
    \qbezier(150,10)(140,0)(130,10)
    \qbezier(130,10)(120,20)(110,10)
	\qbezier(110,10)(100,0)(110,-10)
    \qbezier(-10,0)(-10,-40)(100,-40)
    \qbezier(100,-40)(140,-40)(150,-30)
    \qbezier(150,-30)(160,-20)(150,-10)
    \qbezier(150,-10)(140,0)(130,-10)
    \qbezier(130,-10)(120,-20)(110,-10)
    \qbezier(30,0)(50,10)(70,0)
    \qbezier(30,0)(50,-10)(70,0)
    \qbezier(30,0)(28,1)(26,2.5)
    \qbezier(70,0)(72,1)(74,2.5)
\linethickness{.6pt}
\put(33.0,-29.0){$A_1$}
\put(44.75,-19.80){\circle*{1.5}}
    \qbezier(49.9,-5.5)(39.5,-20.5)(49,-37)
    \put(50.73,-5.5){\circle*{0.8}}
    \put(51.60,-7.1){\circle*{0.8}}
    \put(52.65,-9){\circle*{0.8}}
    \put(53.50,-11){\circle*{0.8}}
    \put(54.15,-13){\circle*{0.8}}
    \put(54.60,-15){\circle*{0.8}}
    \put(54.85,-17){\circle*{0.8}}
    \put(54.93,-19){\circle*{0.8}}
    \put(54.95,-21){\circle*{0.8}}    
    \put(54.90,-23){\circle*{0.8}}
    \put(54.75,-25){\circle*{0.8}}
    \put(54.40,-27){\circle*{0.8}}
    \put(53.85,-29){\circle*{0.8}}
    \put(53.10,-31){\circle*{0.8}}
    \put(52.15,-33){\circle*{0.8}}
    \put(51.00,-35){\circle*{0.8}}
    \put(49.80,-36.6){\circle*{0.8}}

\put(104,-27){$A_2$}
    \qbezier(119.9,-15)(112.9,-26.8)(119.7,-39)
    \put(120.55,-15.2){\circle*{0.8}}
    \put(121.50,-17){\circle*{0.8}}
    \put(122.40,-19){\circle*{0.8}}
    \put(123.05,-21){\circle*{0.8}}
    \put(123.50,-23){\circle*{0.8}}
    \put(123.75,-25){\circle*{0.8}}
    \put(123.82,-27){\circle*{0.8}}  
    \put(123.75,-29){\circle*{0.8}}
    \put(123.50,-31){\circle*{0.8}}
    \put(123.05,-33){\circle*{0.8}}
    \put(122.40,-35){\circle*{0.8}}
    \put(121.50,-37){\circle*{0.8}}
    \put(120.50,-38.8){\circle*{0.8}}

\put(48,23){$B_1$}
    \qbezier(20,0)(20,20)(50,20)
    \qbezier(80,0)(80,20)(50,20)
    \qbezier(20,0)(20,-20)(50,-20)
    \qbezier(80,0)(80,-20)(50,-20)

\put(140,5.1){\circle*{2.3}}
\put(135,12){$\infty_+$}
\put(140,-5.1){\circle*{2.3}}
\put(135,-14){$\infty_-$}
\put(260,0){\makebox(0,0)[cc]
     {\begin{tabular}{cc}
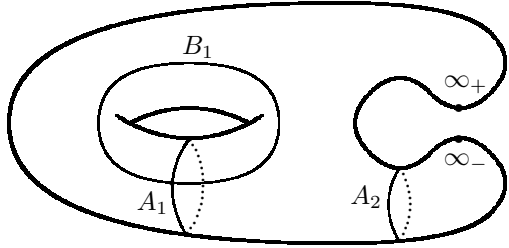
      Fig.2: The canonical homology basis \\
	     of the affine curve C
      \end{tabular}
	    }}
\end{picture}
\end{center}
\vspace{2cm}

\noindent
is just the first projection $\phi(z_1,z_2)=z_1$, and as
$$
	\phi(\Lambda _2)
      = \int_{A _2}\frac{d\lambda}{\mu}
      = 0
$$
then $\phi(\Lambda )$ is generated by $ \phi(\Lambda_1)$ and
$\phi(\Lambda_3)$, and
$$
	 {\rm Ker} \, \phi
       = \CC \Big/ \Big\{
	 \ZZ \int_{A_2}\frac{\lambda \, d\lambda} {\mu}
		   \Big\}
    \sim \CC^* \; .
$$
 As an analytic manifold the usual Jacobian $J(C)$ is
$$
       \CC / \phi(\Lambda) \ \sim \
	H^0 ( C , \Omega^1) ^*
	    \, \big/ \, H_1(C,\ZZ) \; .
$$

In contrast to the usual Jacobian  $J(C)$, the generalized
Jacobian    
$\CC^2/\Lambda$ is a {\it non--compact} algebraic group.
For any $p\in J(C)$ define also the  divisor
$D_p=\phi^{-1}(p)\subset J(C;\infty^\pm)$.

    An explicit embedding of a Zariski open subset of $J(C;\infty^\pm)$
in $\CC^6$ is constructed by the following classical construction
due to Jacobi  (see Mumford\cite{Mumford}).
Let
\begin{equation}
  \label{polynom}
    f(\lambda)
      = \lambda^4 + a_1 \lambda^3 + a_2 \lambda^2
		  + a_3 \lambda + a_4
\end{equation}
be a polynomial without double roots and define the polynomials
\begin{equation}
\label{jp}
	U(\lambda) = \lambda^2	 + u_1 \lambda + u_2 \ , \qquad
	V(\lambda) = v_1 \lambda + v_2		     \ , \qquad
	W(\lambda) = \lambda^2	 + w_1 \lambda + w_2 \ .
\end{equation}
Let $T_C$ be the set of Jacobi polynomials (\ref{jp}) satisfying the
relation
\begin{equation}
  \label{jp1}
      f(\lambda)-V^2(\lambda) = U(\lambda) W(\lambda) \; .
\end{equation}
More explicitly, let us expand
\\ \vspace{-5mm}
\ben
     f - V^2 - UW
   = \sum_{i=0}^3 b_i(u_1,u_2,v_1,v_2,w_1,w_2) \,
     \lambda^{i}			       \ \ ,
\een
\vmmm \vmmm
\begin{align}
 \nonumber   b_3 & =  a_1 - u_1 - w_1			 &
	     b_2 & =  a_2 - u_2 - w_2 - u_1 w_1 - v_1^2  \\
\nonumber    b_1 & =  a_3 - u_1 w_2 - u_2 w_1 - 2 v_1 v_2 &
	     b_0 & =  a_4 - u_2 w_2 - v_2^2		 \ .
\end{align}
If we take $u_i, v_j, w_k$ as coordinates in $\CC^6$ then $T_C$ is
just the zero locus $V(b_0,b_1,b_2,b_3)$ as a subset of $\CC^6$
\begin{eqnarray*}
	 T_C = \Big\{ (u,v,w)\in \CC^6 & : &
		      u_1 + w_1 = a_1 \,, \ \
		      u_2 + w_2 + u_1 w_1 + v_1^2 = a_2 \,,
\\
	 & &	u_1 w_2 + u_2 w_1 + 2 v_1 v_2  = a_3\,, \
		u_2 w_2 + v_2^2  =  a_4  \Big\} 	\ .
\end{eqnarray*}

\begin{pr}
  \label{pr.algebraic}
      If $f(\lambda)$ is a polynomial without double roots then

    {\rm i)}
       $\ T_C$ is a smooth affine variety  isomorphic to 
       $J(C;\infty^\pm) \setminus D_p \,$ for some $p\in J(C)$

    {\rm ii)}
       any translation invariant vector field  on the
       generalized Jacobian $J(C;\infty^\pm)$ of the curve
       $C$ can be written {\rm (}up to multiplication by
       a non--zero constant{\rm )} in the following Lax
       pair form
\vmm
\begin{equation}
  \label{Lax1}
       2 \sqrt{-1} \, \frac{d}{d\,t} A(\lambda)
     = \biggl[ \,
	    A(\lambda) \, , \, \frac{A(a)}{\lambda - a \, }
       \biggr]
\end{equation}

\vspace{-0.3cm}

\noindent
where
\begin{equation}
  \label{AA}
       A(\lambda) =
    \left(
      \begin{array}{lr}
	  V(\lambda)   &  U(\lambda)   \\
	  W(\lambda)   & -V(\lambda)
      \end{array}
    \right)
\end{equation}
$a\in \CC$, and $U,V,W$ are the Jacobi polynomials {\rm (\ref{jp})}.

      Equivalently, if $D=P_1+P_2 \in {\rm Div}^2
(\breve{C})$, where $P_i=(\lambda_i,\mu_i)$, $i=1,2$,
then {\rm (\ref{Lax1})} can be written as
\vmm
\begin{eqnarray}
	 \frac{d\lambda_1}{\sqrt{f(\lambda_1)}}
       + \frac{d\lambda_2}{\sqrt{f(\lambda_2)}}
    &=& -\sqrt{-1} \, dt
\nonumber \\[-4pt]
    & & \label{abel1} \\[-4pt]
	\frac{\lambda_1 d\lambda_1}{\sqrt{f(\lambda_1)}}
      + \frac{\lambda_2 d\lambda_2}
       {\sqrt{f(\lambda_2)}}
    &=& -a \sqrt{-1} \, dt \nonumber \ \ .
\end{eqnarray}
\end{pr}
{\bf Remark.} Note that $a = \infty $ also makes a sense. The
corresponding vector field is obtained by changing the
time as $t \rightarrow t/a$ and letting $a\rightarrow \infty$.
Thus (\ref{Lax1}) becomes
\begin{equation}
  \label{Lax2}
      2 \sqrt{-1} \, \frac{d}{dt}A(\lambda)
   =  \bigl[ \, A(\lambda) \,,\, A_\infty \,\bigr], \qquad
       A_\infty  =  \left(
	     \begin{array}{rr}
		0   &  -1     \\
	       -1   &	0
	     \end{array}
		     \right) \; .
\end{equation}
\vspace{-0.3cm}
and (\ref{abel1})
\vmmm
\begin{eqnarray}
	    \frac{d\lambda_1}{\sqrt{f(\lambda_1)}}
	  + \frac{d\lambda_1}{\sqrt{f(\lambda_1)}}
      &=& 0
\nonumber
\\[-5pt]
      & & \label{abel2}
\\[-5pt]
	    \frac{\lambda_1 d\lambda_1}{\sqrt{f(\lambda_1)}}
	  + \frac{\lambda_2 d\lambda_2}
	    {\sqrt{f(\lambda_2)}}
      &=& - \sqrt{-1} \, dt
\nonumber \ .
\end{eqnarray}

	 The proof of part i) of the above Proposition can be found
in Previato \cite{Previato} (see also Mumford \cite{Mumford}).
It is also proved there that a translation invariant vector
field $\frac{d}{d \epsilon}$ on the generalized Jacobian
$J(C;\infty^\pm)$ which is induced by the tangent vector
\begin{equation}
\label{ll}
    \frac{d}{d \epsilon} \, \lambda_{\big\lvert \lambda = a}
	   = \sqrt{f(a)}
\end{equation}
on $C$ via the Abel map $C\rightarrow J(C;\infty^\pm)$, can be
written as
\be
\label{uu}
	  \frac{d}{d\epsilon}U(\lambda)
     &=&  \ \ \
	  \frac{V(a) U(\lambda) - U(a)V(\lambda)}{\lambda-a}
\\[8pt] \label{ww}
	  \frac{d}{d \epsilon}W(\lambda)
     &=& -\frac{V(a)W(\lambda) - W(a)V(\lambda)}{\lambda-a}
\\[8pt] \label{vv}
	  \frac{d}{d \epsilon}V(\lambda)
     &=&  \ \ \
	  \frac{U(a)W(\lambda) - W(a)U(\lambda)}
	       {2(\lambda-a)}			     \ .
\ee

       Our final remark is that the translation invariant vector
fields (\ref{uu}), (\ref{ww}), and (\ref{vv}), which we
denote further by $\frac{d}{d t}$,  can be written in the
following Lax pair form (suggested by
Beauville \cite[example 1.5]{Beauville})
$$
   -2 \, \frac{d}{dt} A(\lambda)
       = \biggl[ \,  A(\lambda) 	   \,,\,
		    \frac{A(a)}{\lambda-a} \,
	 \biggr]
$$
where
$$
       A(\lambda)
    = \left( \begin{array}{lr}
		V(\lambda)  &  U(\lambda) \\
		W(\lambda)  & -V(\lambda)
	     \end{array}
      \right) \; .
$$
By (\ref{ll}) the direction of the constant tangent vector
computed above is
$$
       \left(
	   \frac{  \bp{\lambda}}{\sqrt{f(a)}} \ , \
	   \frac{a \bp{\lambda}}{\sqrt{f(a)}}
       \right)
     = (1,a)
$$
which proves (\ref{abel1}). This completes the proof
of Proposition \ref{pr.algebraic}.
{\hfill $\Box $}

\mm

      Next we apply Proposition \ref{pr.algebraic}
to the Lagrange top (\ref{Ltop}). Let  $C_h$ be the curve
$C$ as above, where
\vmm
\begin{equation}
  \label{eah}			      \qquad
      a_1 =  2 (1+m) h_4	   \,,\quad
      a_2 =  2 h_3 + m(m+1) h_4^2  \,,\quad
      a_3 = -2 h_2		   \,,\quad
      a_4 =  h_1  =  1		   \ ,
\end{equation}

\vspace{-0.2cm}
\noindent
so
     \vspace{-0.2cm}
\begin{equation}
\label{fh}
       \breve{C}_h = \Bigl\{ \, \mu^2
     = \lambda^4 + 2(1+m)h_4 \lambda^3
	 + \bigl( 2h_3 + m(m+1)h_4^2 \bigr) \lambda^2
	 -  2h_2 \lambda + 1 \, \Bigr\} \; .
\end{equation}
\vmmm
Consider  the complex invariant level set of the
Lagrange top (\ref{Ltop})
\mm
\ben
     T_h = \Bigl\{ \, (\Omega ,\Gamma)\in \CC^6:
		    H_1 (\Omega ,\Gamma) = 1,	 \,
		    H_2 (\Omega ,\Gamma) = h_2,  \,
		    H_3 (\Omega ,\Gamma) = h_3,  \,
		    H_4 (\Omega ,\Gamma) = h_4	 \,
	    \Bigr\}
\een
and the associated ``bifurcation set''
$$
       {\bf B}
    =  \bigl\{
	  h\in \CC^3: {\rm discriminant} \,
	  \bigl( f(\lambda) \bigr)
    = 0
       \bigr\} \ .
$$
It is a straightforward computation to check that
the linear change of variables
\begin{align}	  \qquad
   u_1 & = (1+m)\Omega_3  - i \Omega_2	   &
   u_2 & =  -\Gamma_3	  + i \Gamma_2	   &
			  \nonumber
\\		   \qquad \label{change}
   w_1 & =  (1+m)\Omega_3 + i \Omega_2	   &
   w_2 & =  - \Gamma_3	  - i \Gamma_2	   &
\\		   \qquad  \nonumber
   v_1 & =  \Omega _1			   &
   v_2 & =  - \Gamma_1	\qquad	\qquad
		    \qquad i = \sqrt{-1} \ &
\end{align}
identifies $T_C$ and $T_h$. Further, as
$$
    \biggl[  A(\lambda) \,,\, \frac{A(a)}{\lambda -a}
    \biggr]
  = \biggl[  A(\lambda) \,,\,
	     \frac{A(a) - A(\lambda )}{\lambda -a}
    \biggr]
  = \biggl[  A(\lambda),\left(
		     \begin{array}{cc}
		      -v_1	  & - a\! - \!u_1\! - \!
				     \lambda \\
   - a\! - \!w_1\! - \!\lambda	&   v_1
		     \end{array} \right)
    \biggr]
$$
then the vector field (\ref{Ltop}) is obtained by
substituting $a = -m \Omega_3$ in (\ref{Lax1}) and
using the change of variables (\ref{change}) (note
that $\Omega_3$ is a constant of motion). Similarly
the vector field (\ref{Ltop1}) is obtained
by substituting $a = \infty$ (see the remark after
Proposition \ref{pr.algebraic}).

    To summarize we proved the following

\begin{th1}
  \label{th.Lagrange}
If $h\not\in {\bf B}$ then

  {\rm i)}
      the complex invariant level set $T_h$ of the Lagrange
      top is a smooth complex manifold	biholomorphic to
      $J(C_h;\infty^\pm) \setminus D_\infty$ where
      $D_\infty = \phi^{-1}(p)$ for some $p\in J(C_h)$
      and $J(C_h;\infty^\pm)$ is the generalized
      Jacobian of the elliptic curve $C_h$ with two points
      at ``infinity'' identified.

  {\rm ii)} The  Hamiltonian flows of the Lagrange top
      {\rm (\ref{Ltop}), (\ref{Ltop1})} restricted to
      $T_h$ induce linear flows on $J(C_h;\infty^\pm)$. The
      corresponding vector fields {\rm (\ref{Ltop})} and
      {\rm (\ref{Ltop1})} have a Lax pair representation
      obtained from the Lax pair {\rm (\ref{Lax1})} by
      substituting $a = -m \Omega_3$ and $a = \infty$
      respectively, and using the change of variables
      {\rm (\ref{change})}.
\end{th1}

      According to the above theorem the Lagrange top
is an algebraically completely integrable system in the
sense of Mumford \cite[p.353]{Mumford}. Clearly any
linear flow on $J(C_h;\infty^\pm)$ maps under $\phi$
(\ref{extension}) into a linear flow on the usual
Jacobian $J(C_h)$. This is expressed by the fact that
the variable $\Gamma_3$ which describes the nutation
of the body is an elliptic function in time. It was
known to Lagrange \cite{Lagrange} who deduced the
differential equation satisfied by  $\Gamma_3$. The real
version of Theorem \ref{th.Lagrange} will be explained in
section \ref{real}.

      To the end of this section we compare the Lax
pair (\ref{Lax1}) and the Lax pair for the Lagrange top
obtained earlier by Adler and van Moerbeke \cite{Adler}.
Namely, if we identify the Lie algebras $(\RR^3, \wedge)$
and $\bigl( {\rm so(3)},[\,.\,,\,.\,] \bigr) $ by the Lie
algebras isomorphism
$$
      \left(
	 \begin{array}{c}
	      x \\
	      y \\
	      z
	 \end{array}
      \right)
    \rightarrow
      \left(
	 \begin{array}{ccc}
	     0	& -z &	y  \\
	     z	&  0 & -x  \\
	    -y	&  x &	0
	 \end{array}
     \right)		       \ \,,
$$
then (\ref{Ltop}) can be written in the following
equivalent Lax pair form \cite{Adler}
\vmm
\begin{equation}
\label{LaxL}
      \frac{d}{dt} \,
      \big( \lambda^2 \chi + \lambda M -\Gamma \big)
  = \big[ \,\lambda^2 \chi + \lambda M - \Gamma \,,\,
       \lambda \chi+ \Omega \, \big]		\ .
\end{equation}
\vspace{-0.1cm}
where
\vmm
$$
  \Omega = (\Omega_1 , \Omega_2 , \Omega_3),	     \
       M = \big( \Omega_1 , \Omega_2 , (1+m)\Omega _3 \big) ,\
  \Gamma = (\Gamma_1 , \Gamma_2 , \Gamma_3),	     \
    \chi = (0,0,1)				  \; .
$$
The Lax pair representation of (\ref{Ltop1}) is given by
\begin{equation}
\label{LaxA+}
   \frac{d}{dt} \,\big(\lambda^2 \chi+\lambda M -\Gamma\big)
	 = \big[\,\lambda^2 \chi + \lambda M - \Gamma \,,\,
	      \chi \, \big] \ .
\end{equation}
Both Lax pairs (\ref{LaxL}), (\ref{LaxA+}) can be also written in
the Beauville form
\begin{equation}
\label{LaxB}
       - \frac{d}{dt}A(\lambda)
     = \biggl[	  \, A(\lambda)\,,\,\frac{A(a)}{\lambda - a}
       \biggr]
\end{equation}
where $A(\lambda)= \lambda^2 \chi+\lambda M -\Gamma$. Indeed,
$$
      \biggl[ \, A(\lambda)\,,\,\frac{A(a)}{\lambda-a}	 \, \biggr]
   =  \biggl[ \, A(\lambda)\,,\,\frac{A(a)-A(\lambda)}{\lambda-a}
							    \biggr]
   = -\biggl[ \, A(\lambda)\,,\,\lambda \chi + a \chi + M \,\biggr] \ .
$$
Now (\ref{LaxL}) is obtained by replacing as before $a=-m \Omega_3$,
and (\ref{LaxA+}) is obtained by letting $a\rightarrow \infty$.

       Clearly the Lax pair (\ref{Lax1}) from Proposition
\ref{pr.algebraic} and (\ref{LaxL}), (\ref{LaxB}) are
equivalent in the sense that they define one and the same
vector field. We can identify them over $\CC$ by making
use of the isomorphism of the Lie algebras
${\rm so}(3,\CC)$ and ${\rm sl}(2,\CC)$ given by
$$
       \left(
	  \begin{array}{ccc}
	     0	& -z  &   y   \\
	     z	&  0  &  -x   \\
	    -y	&  x  &   0
	  \end{array}
       \right)
    \rightarrow
      \frac{1}{\sqrt{2}}
      \left(
	  \begin{array}{cc}
	    \epsilon x & \epsilon z + \overline{\epsilon} y  \\
	    \epsilon z - \overline{\epsilon} y & -\epsilon x
	  \end{array}
      \right) ,
  \qquad
     \epsilon = \exp{\frac{\sqrt{-1} \pi}{4}} \; .
$$
Note, however, the following difference. The spectral
curve of (\ref{LaxL}) is reducible
$$
   {\rm det} \,
 \big( \, \lambda^2 \chi + \lambda M + \Gamma - \mu I \, \big)
   = -\mu \, \bigl( \mu^2 + f(\lambda) \bigr) = 0     \ ,
$$
$$
	f(\lambda)  =  \lambda^4 + 2(1+m)h_4 \lambda^3
    +  \bigl( 2h_3 + m(m+1)h_4^2 \bigr) \lambda^2
    -  2 h_2 \lambda + 1     \ ,
$$
but the spectral curve of (\ref{Lax1}) is not
$$
      {\rm det} \bigl(
		  A(\lambda) - \mu I \bigr)
		= \mu^2 - V^2 - U W
		= \mu^2 - f(\lambda)
		=   0 \ .
$$
The last observation will be of some importance for the next
section. Earlier Adler and van Moerbeke \cite[p.351]{Adler}
proposed to linearize the Lagrange top on an elliptic curve by
introducing first a small parameter $\epsilon$ in the
corresponding ${\rm so}(3)$ Lax pair. The new system has the
advantage to have an irreducible genus four spectral
curve $C_\epsilon$ which fits to the general theory,
so we can just ``take the limit''
$\epsilon \rightarrow 0$.  This computation reproduced in
\cite{Ratiu} and used in \cite{Ratiu1} is however erroneous.

    By abuse of notation we call the curve
$\tilde{C}_h = \{ \mu^2 + f(\lambda) = 0 \}$  with an
antiholomorphic involution
$(\lambda, \mu) \rightarrow (\overline{\lambda},\overline{\mu})$,
{\it the spectral curve} of the Lax pair (\ref{LaxL}). The curve
$\tilde{C}_h$ is real isomorphic to the curve
$C_h = \bigl\{ \mu^2 = f(\lambda) \bigr\}$,
equipped with an antiholomorphic involution
$(\lambda, \mu) \rightarrow (\overline{\lambda},- \overline{\mu})$,
so without loss of generality we shall write $\tilde{C}_h = C_h $.


\section{Explicit Solutions}
\label{solutions}

      In this section we find explicit solutions for the Lagrange
top (\ref{Ltop}). We compute first the Baker--Akhiezer
function of the ${\rm sl}(2,\CC)$ (or rather ${\rm su(2)}$)
Lax pair
(\ref{Lax1}). This implies explicit formulae for the solutions
of the Lagrange top in terms of exponentials and theta functions
related to the spectral curve $C_h$ (see for example Dubrovin
\cite{Dubrovin3}, E.D.Belokolos, A.I.Bobenko, V.Z.Enol'skii,
A.R.Its, V.B.Matveev \cite{Belokolos}).Then we note that
the Jacobian $J(C_h)$ of $C_h$ is just	the Lagrange
elliptic curve used in the
classical theory which provides explicit solutions in
terms of exponentials and sigma function related to $J(C_h)$.

       By performing an unitary operation on the matrix (\ref{AA})
we may put its leading term in diagonal form. Substituting
$a=-m \Omega _3$ in (\ref{Lax1}) and using the change of the
variables (\ref{change}) we obtain the following Lax pair
representation for the Lagrange top (\ref{Ltop})
\begin{equation}
\label{LaxA}
       \biggl[ \, A  \,,\,
	 B - 2i \frac{d}{dt} \, \biggr]
    \, = \, 2i \frac{dA}{dt} + [A,B]
    \, = \, 0 \,,
  \quad
    \epsilon^2 = i \,,
  \quad
	  i^2  = -1
\end{equation}
where
\begin{eqnarray*}
    A  & = &  A(t,\lambda )
       =  \left(
	    \begin{array}{lr}
		A_{11}(t,\lambda) & A_{12}(t,\lambda) \\
		A_{21}(t,\lambda) & A_{22}(t,\lambda)
	    \end{array}
	  \right)
       =  \left(
	    \begin{array}{lr}
		1  &  0  \\
		0  & -1
	    \end{array}
	  \right)
	 \lambda^2 \\[6pt]
       & + &
	 \left(
	    \begin{array}{cc}
		 (1+m)\Omega _3 				   &
		 \overline{\epsi}\Omega _1(t) + \epsi \Omega _2(t) \\
		 \epsi \Omega _1(t) + \overline{\epsi}\Omega _2(t) &
		 -(m+1)\Omega _3
	     \end{array}
	  \right)
	  \lambda
	- \left(
	    \begin{array}{cc}
		   \Gamma_3 & \overline{\epsi}\Gamma_1(t)+\epsi
		      \Gamma_2(t) \\
		   \epsi \Gamma_1(t)
		     + \overline{\epsi}\Gamma_2(t)		&
		     -\Gamma_3
	    \end{array}
	  \right)
\end{eqnarray*}
and
$$
    B = B(t,\lambda) =
       \left(
	    \begin{array}{cr}
	       1  &  0	\\
	       0  & -1
	    \end{array}
       \right) \lambda
    + \left(
	    \begin{array}{cc}
		 \Omega _3
	       &  \overline{\epsi} \Omega _1(t)
		  + \epsi  \Omega_2(t) \\
		    \epsi \Omega_1(t)
		  + \overline{\epsi} \Omega_2(t)
	       & - \Omega_3
	   \end{array} \right) \ .
$$
The spectral curve of the above Lax representation is
defined by
$$
	\breve{C}_h
     = \bigl\{ \, {\rm det}
	       \bigl( A(\lambda) - \mu I \bigr)
     = \mu^2 - f(\lambda) = 0 \, \bigr\} \,,
$$
$$
	f(\lambda)
     = \lambda^4 + 2(1+m)h_4 \lambda^3 +
       \bigl( 2h_3 + m (m+1) h_4^2 \bigr) \lambda^2
       - 2 h_2 \lambda + 1  \ .
$$
We shall also denote by $C_h$ the Riemann surface of the
compactified affine curve $\breve{C}_h$. The reader may
note the ``similarity'' between (\ref{LaxA}) and the Lax
pair of the nonlinear Schr\"odinger equation (for a
rigorous statement see Proposition \ref{sch.pr}).

\subsection{The Baker--Akhiezer function}

    Let us fix a solution $A(t,\lambda )$ of (\ref{LaxA}) defined
in a neighborhood of $t=0\in \CC$. We shall also suppose that the
point $P=(\lambda ,\mu )$ is such that $(1,-1)$ is not an
eigenvector of the matrix $A(0,\lambda )$.

\begin{pr}
  \label{Ap1.pr}
For any $t \in \CC$ in a sufficiently small neighborhood
of the origin, there exists an unique eigenfunction
\begin{equation}
  \label{A7}
     \Psi = \Psi (t,P)
	  = \left(
	      \begin{array}{c}
		 \Psi^1 (t,P)	     \\
		 \Psi^2 (t,P)
	      \end{array}
	    \right) ,	 \quad
	P = (\lambda, \mu) \in \breve{C}
\end{equation}
of $A(t,\lambda)$ {\rm (}called Baker--Akhiezer
function{\rm )} satisfying the conditions
\begin{align}
	    2 i \frac{d}{dt} \Psi (t,P)
       =&  \, B(t,\lambda) \Psi (t,P)	       &
  \label{A8}\\[4pt]
	    A(t,\lambda) \Psi (t,P)
       =&  \, \mu \Psi (t,P)		       &
  \label{A9}
\end{align}
and normalized by
\begin{equation}
  \label{A10}
	\Psi^1 (0,P) + \Psi^2 (0,P) = 1 \ .
\end{equation}
In terms of the coefficients $A_{ij}(t,\lambda)$ of the matrix
$A = (A_{ij})$ we have
\be		\
  \label{A11}
	 \Psi^1(0,P)
       = \frac{A_{12}(0,\lambda)}
	      {A_{12}(0,\lambda) + \mu - A_{11}(0,\lambda)}
       = \frac{\mu - A_{22}(0,\lambda)}
	      {A_{21}(0,\lambda) + \mu - A_{22}(0,\lambda)} \ \
\\[8pt]  \label{A11+}  \
	 \Psi^2(0,P)
       = \frac{\mu - A_{11}(0,\lambda)}
	      {A_{12}(0,\lambda) + \mu - A_{11}(0,\lambda)}
       = \frac{A_{21}(0,\lambda)}
	      {A_{21}(0,\lambda) + \mu - A_{22}(0,\lambda)} \ .
\ee
\end{pr}
\noindent
{\bf{Proof.}}
Let $\Phi(t,\lambda)$ be a fundamental matrix for the operator
$B(t,\lambda)-2 i \frac{d}{dt}$ normalized at $t=0$. Then the
general solution of (\ref{A8}) is written as
\begin{equation}
\label{A12}
	  \Psi(t,P) = \Phi(t,\lambda) \Psi(0,P),  \quad
    \Phi(0,\lambda) = \left(
			 \begin{array}{lr}
			     1	&  0  \\
			     0	&  1
			 \end{array} \right),	  \quad
		  P = (\lambda,\mu) \; .
\end{equation}
As $A$ and $B-2 i \frac{d}{dt}$ commute then
$$
   \Bigl( B(t,\lambda) - 2i \frac{d}{dt} \Bigr)  \,
	  A(t,\lambda) \Phi(t,\lambda)
    \, = \,
	  A(t,\lambda)
   \Bigl( B(t,\lambda) - 2i \frac{d}{dt} \Bigr)  \,
	  \Phi(t,\lambda)
    \, = \, 0
$$
and hence
$A(t,\lambda) \Phi(t,\lambda) = \Phi(t,\lambda) M(P)$
for some constant matrix $M(P)$ computed by substituting
$t=0$. Thus $M(P)=A(0,\lambda)$ and
$$
	 A(0,\lambda)
      = \Phi^{-1}(t,\lambda) A(t,\lambda) \Phi(t,\lambda) \ .
$$
The constants $\Psi^1(0,P), \Psi^2(0,P)$ are uniquely defined by
(\ref{A9}) and (\ref{A10}). Finally
\begin{eqnarray*}
	  A(t,\lambda) \Psi(t,P )
     &=&  \Phi(t, \lambda )  A(0,\lambda)
	  \Phi^{-1}(t,\lambda ) \Phi(t,\lambda ) \Psi(0,P)
\\
     &=&  \Phi(t,\lambda ) .\mu . \Psi(0,P)
\\
     &=&  \mu \Psi(t,P) \; .
\end{eqnarray*}
The formulae (\ref{A11}), (\ref{A11+}) follow from  (\ref{A9}),
(\ref{A10}). {\hfill $\Box $}

\mm

      Denote by $\infty^+$ (respectively $\infty^-$) the point on
$C_h - \breve{C}_h$ such that in its neighborhood \-
$\mu/\lambda^2 \sim + 1$ (resp. $(-1)$).
\begin{pr}
  \label{Ap2.pr}
There exists $t_0>0$ such that for any fixed
$t\in \CC$, $|t| < t_0$, the Baker--Akhiezer vector--function
$\Psi(t,P)$ is meromorphic in $P$ on the affine curve
$\breve{C_h}$ and it has two  poles at $P_1, P_2 \in C_h$ which
do not depend on $t$. In a neighborhood of the two infinite
points $\infty^{\pm}$ on $C_h$ we have
\begin{eqnarray}		   \ \
	  \Psi^1(t,P)
    & = & \left \{
	   \begin{array}{rll}
	      \bigl( 1 + O(\lambda^{-1}) \bigr) 	 &
	      \!\!\!\!\exp \bigl( -\frac{i}{2}
	      (\lambda + \Omega_3)t \bigr) ,		 &
	      P \rightarrow \infty^+,
	      \ \ i = \sqrt{-1}   \\[2pt]
	      O(\lambda^{-1})				 &
	      \!\!\!\!\exp \bigl( + \frac{i}{2}
	      (\lambda + \Omega_3)t \bigr) ,		 &
	      P \rightarrow \infty^-
	   \end{array}	\right.
  \label{A1}\\[4pt]		   \ \
	   \Psi^2(t,P)
    & = & \left\{
	    \begin{array}{rll}
		 O(\lambda^{-1})			 &
	       \!\!\!\!\exp \bigl( -\frac{i}{2}
	       (\lambda + \Omega_3)t \bigr) ,		 &
	       P \rightarrow \infty^+			\\[2pt]
	       \bigl( 1 + O(\lambda)^{-1} \bigr)	 &
	       \!\!\!\! \exp \bigl( + \frac{i}{2}
	       (\lambda +\Omega_3)t \bigr) ,		 &
	       P \rightarrow \infty^-
	    \end{array}  \right.
  \label{A2}
\end{eqnarray}
Moreover, $\Psi^1(t,P)$ $\,{\rm (} \,\Psi^2(t,P)\, {\rm )}$
 has exactly one zero
on $\breve{C}_h$ and the refined asymptotic estimates of
$\Psi^1$ at $\infty^-$ and  of $\Psi^2$ at $\infty^+$ read
\begin{eqnarray}	       \qquad
       \Psi^1(t,P) & = &
       \biggl[	- \frac{\overline{\epsi} \Omega_1(t)
		+ \epsi \Omega_2(t)} {2 \lambda}
		+ O({\lambda^{-2}})
       \biggr]
       \exp \bigl( + \tfrac{i}{2}(\lambda + \Omega_3)t
	    \bigr), \
       P \rightarrow \infty^-
  \label{e18a}\\	       \qquad
       \Psi^2(t,P) & = &
       \biggl[	+ \frac{\epsi \Omega_1(t)
		+ \overline{\epsi} \Omega_2(t)}{2 \lambda}
		+ O(\lambda^{-2})
       \biggr]
       \exp \bigl( - \tfrac{i}{2} (\lambda + \Omega_3)t
	    \bigr), \
       P \rightarrow \infty^+ \ .
  \label{e18b}
\end{eqnarray}
\end{pr}

\noindent
{\bf {Proof.}}
According to (\ref{A9}) $(\Psi^1,\Psi^2) \in {\rm Ker}(A-\mu I)$
and hence
\begin{equation}
  \label{p1p2}
	 \frac{\Psi^2(t,P)}{\Psi^1(t,P)}
       = \frac{ \mu - \lambda^2
	       - (1+m)\Omega_3 \lambda + \Gamma_3(t)}
	      { \bigl( \overline{\epsi} \Omega_1(t)
	       + \epsi \Omega_2(t) \bigr) \lambda
	       - \overline{\epsi} \Gamma_1(t)
	       + \epsi \Gamma_2(t)} \ .
\end{equation}

      If $P \rightarrow \infty^+$ then
$\mu - \lambda^2 - (1+m)\Omega_3 \lambda \sim O(1)$ and using
(\ref{LaxA}), (\ref{A8}), (\ref{A9}) and (\ref{p1p2}) we compute
$$
	 2i \frac{d}{dt} \ln \Psi^1(t,P)
       = \lambda + \Omega_3
       + \bigl( \overline{\epsi} \Omega_1(t)
	      + \epsi \Omega_2(t) \bigr)
	 \frac{\Psi^2(t,P)}{\Psi^1(t,P)}
       = \lambda + \Omega_3 + O(\lambda^{-1})
$$
and hence
$$
	\Psi^1(t,P)
      = \bigl( 1 + O(\lambda^{-1} \bigr) \,
	\exp \bigl( - \tfrac{i}{2}(\lambda + \Omega_3)t
	     \bigr)  \ .
$$
In a similar way if $P \rightarrow \infty^-$ we obtain
$$
	\Psi^2(t,P)
     =	\bigl( 1 + O(\lambda^{-1}) \bigr) \,
	\exp \bigl( + \tfrac{i}{2} (\lambda + \Omega_3) t
	     \bigr) \ .
$$
To compute the remaining asymptotic estimates we use
that if $P \rightarrow \infty^-$  then
\begin{equation}
  \label{e15}
	  \frac {\Psi^1(t,P)} {\Psi^2(t,P)}
      =   \frac {A_{12}(t,\lambda )} {\mu - A_{11}(t,\lambda )}
      = - \frac {\overline{\epsi} \Omega_1(t)
		  + \epsi \Omega_2(t)}
		{2 \lambda } + O(\lambda^{-2})
\end{equation}
and if $P \rightarrow \infty^+$ then
\begin{equation}
  \label{e16}
	 \frac{\Psi^2(t,P)}{\Psi^1(t,P)}
       = \frac{A_{21}(t,\lambda )}{\mu - A_{22}(t,\lambda )}
       = \frac{\epsi \Omega_1(t) + \overline{\epsi}
	       \Omega_2(t)} {2\lambda} + O(\lambda^{-2}) \ .
\end{equation}

      To find the poles of $\Psi(t,P)$ in $P$ we note that
according to the proof of Proposition \ref{Ap1.pr} (and with
the same notations) we have
\begin{equation}
  \label{psi}
      \Psi(t,P)       = \Phi(t,\lambda) \Psi(0,P),  \quad
      \Phi(0,\lambda) = I_2 \; .
\end{equation}
If $|t|$ is sufficiently small, the fundamental matrix
$\Phi(t,\lambda)$ has no poles and ${\rm det}\,
\Phi(t,\lambda) \neq 0$.
It follows that the poles of $\Phi(t,\lambda)$ and
$\Phi(0,\lambda)$ coincide, and we can obtain them
by solving the following quadratic equation
\ben
	{\rm det} \, A(0,\lambda)
     =	\bigl( A_{11}(0,\lambda) - A_{12}(0,\lambda)
	\bigr) ^2
     = \mu^2
\een
$\bigl($see (\ref{LaxA}, (\ref{A11})$\bigr)$.
One gets two time independent
poles $P_1, P_2 \in \breve{C}_h$ of $\Psi(t,P)$.

      At last the meromorphic one--form $d \ln \Psi^1$ has a simple
pole at $\infty^-$ with residue $+1$ and is holomorphic in a
neighborhood of $\infty^+$. On the other hand $\Psi^1(t,P)$ has
exactly two poles on $\breve{C}_h$ and hence it has one zero on
$\breve{C}_h$. The same arguments hold for $\Psi^2(t,P)$.
{\hfill $\Box $}

\mm

     Let $A_1, A_2, B_1$ be a basis  of $H_1(\breve{C_h},\ZZ)$ as
it is shown on figure 2 ($A_1 \circ B_1 = 1)$,
$\omega_1$, $\omega_2$ be a basis of
$H^0 \big( C , \Omega^1 (\infty^+ + \infty^- ) \big)$,
normalized by the conditions
$$
    \left(
	  \int_{A_i} \omega_j
    \right)_{i,j=1,2}
  = \left(
	    \begin{array}{cc}
		2\pi i &  0	  \\
		 0     &  2\pi i
	    \end{array}
    \right) \; .
$$
We shall also suppose that $\omega_1$ is a holomorphic form
on the elliptic curve $C_h$. Define now the period matrix
$$
	\Pi = \left(
		    \begin{array}{lcr}
		      2 \pi i  &    0	  &  \tau_1   \\
			 0     &  2\pi i  &  \tau_2
		    \end{array}
								    \right)
$$
where
$$
       \tau_1 = \int_{B_1} \omega_1 \ ,   \qquad
       \tau_2  = \int_{B_1} \omega_2 \ ,   \qquad
       {\rm Re \,}(\tau _1)  <	0		  \ .
$$
Recall that the generalized Jacobian $J(C_h ; \infty^\pm)$ of
$C_h$ relative to the modulus  $m=\infty^++\infty^-$ is
identified with $\CC^2/\Lambda $ where $\Lambda $ is the lattice
in $\CC^2$ generated by the columns of $\Pi $. Let
\vmm
$$
       \theta_{11} (z)
     = \theta_{11} (z \, | \, \tau_1)
     = \sum_{n = - \infty}^{\infty}
	   \exp \Bigl\{ \tfrac{1}{2} \tau_1
	      (n + \tfrac{1}{2})^2
		 + (z + \pi \sqrt{-1})(n + \tfrac{1}{2})
		 \Bigr\},
	 \quad
     z \in \CC
$$
\vspace{-10mm}
$$ $$
be the Jacobi theta function with characteristics
$\left[ \tfrac{1}{2} \,,\,\tfrac{1}{2} \right] $,
$$
      \theta_{11}(0)	    =  0, \quad
      \theta_{11}(z+2\pi i) =  -\theta_{11}(z),  \quad
      \theta_{11}(z+\tau_1) = - \exp \big( -z-\tfrac{1}{2} \tau_1 \big)
				\,     \theta _{11}(z).
$$

     Denote by $\Omega	$  the unique Abelian differential of
second kind on $C_h$ with poles at $\infty^{\pm}$, principal
parts $\pm \frac{i}{2}\, d\lambda $ where $P = (\lambda ,\mu )$,
$i=\sqrt{-1}$, and normalized by $\int_{A _1}\Omega  =0$. Let
$P_0 \in \breve{C}_h$ be a fixed initial point, $c^\pm$, $U$
be the constants defined by
\begin{equation}
  \label{ccu}
      \int_{P_0}^P\Omega
    = \left\{	  \begin{array}{lr}
		     - \frac{i}{2}  \lambda
		     + c^- + 0(\lambda^{-1}) \ ,  &\ \
		       P \rightarrow \infty^+
		    \\[4pt]
		     + \frac{i}{2}  \lambda
		     + c^+ + 0(\lambda^{-1} ) \ , &\ \
			 P \rightarrow \infty^-
		  \end{array}
      \right. , 		\qquad
      \int_{B_1} \Omega = U	\ .
\end{equation}
Define the Abel--Jacobi map
$$
	 {\cal A} \, : \, {\rm Div}^0(C_h) \rightarrow	J(C_h)
		  \  : \  \sum P_i - \sum Q_i  \
     \mapsto				       \
	 \int_{\sum Q_i}^{\sum P_i} \omega_1 \; .
$$
Here, and henceforth, we make the convention that the paths of
integration between divisors are taken within $C_h$ cut along
its homology basis $A_1$, $B_1$, which we assume that
not contain points of these divisors.

\begin{pr}
  \label{Ap5.pr}
The Baker--Akhiezer function is explicitly given by
\mm
\be
  \label{22a}  \qquad
	 \Psi^1(t,P) = {\rm const}_1 \,
		       \exp \Bigl[
		     t \bigl( \int_{P_0}^P \Omega
		     - c^-  - \tfrac{i}{2} \Omega_3
		       \bigr) \Bigr]
	 \frac{\theta_{11}\bigl({\cal A}
	       ( P + \infty^- - P_1 - P_2) + tU \bigr) }
	      {\theta_{11} \bigl( {\cal A}
			   (\infty^+ +\infty^- - P_1 - P_2) + tU
			   \bigr) }
  \\[8pt]  \label{22b}	 \qquad
	  \Psi^2(t,P) = {\rm const}_2 \,
			\exp \Bigl[
		      t \bigl( \int_{P_0}^P \Omega
		      - c^+ + \tfrac{i}{2} \Omega_3
			\bigr) \Bigr]
	  \frac{\theta_{11}\bigl({\cal A}
		( P + \infty^+ - P_1 - P_2) + tU \bigr) }
	       {\theta_{11} \bigl( {\cal A}
			    (\infty^+ +\infty^- - P_1 - P_2) + tU
			    \bigr) }
\ee
where
\ben
      {\rm const}_1
    = \frac{\theta_{11}\bigl( {\cal A} (P-\infty^-)
		       \bigr) }
	   {\theta_{11}\bigl( {\cal A} (\infty^+-\infty^-)
		       \bigr) }  \cdot
      \frac{\theta_{11}\bigl( {\cal A} (\infty^+-P_1)
		       \bigr) }
	   {\theta_{11}\bigl( {\cal A} (P-P_1)
		       \bigr) }  \cdot
      \frac{\theta_{11}\bigl( {\cal A} (\infty^+-P_2)
		       \bigr) }
	   {\theta_{11}\bigl( {\cal A}(P-P_2)
		       \bigr) }
 \\[8pt]
      {\rm const}_2
    = \frac{\theta_{11} \bigl( {\cal A} (P - \infty^+)
			\bigr) }
	   {\theta_{11} \bigl( {\cal A} (\infty^- - \infty^+)
			\bigr) }  \cdot
      \frac{\theta_{11} \bigl( {\cal A} (\infty^- - P_1)
			\bigr) }
	   {\theta_{11} \bigl( {\cal A}(P - P_1)
			\bigr) }  \cdot
      \frac{\theta_{11} \bigl( {\cal A} (\infty^- -P_2)
			\bigr) }
	   {\theta_{11} \bigl( {\cal A}(P - P_2)
			\bigr) }
\een
and $P_1$, $P_2$ are the poles of $\Psi $.
\end{pr}

      The proof of the above proposition is based on a general
fact: the properties of $\Psi$ enumerated in Proposition
\ref{Ap2.pr} define it uniquely. Indeed, if $\Psi$ and
$\tilde{\Psi}$ are vector functions both satisfying the assertions
of Proposition \ref{Ap2.pr}, then the meromorphic on $C_h$
functions  $\Psi^1$ and $\tilde{\Psi}^1$ (resp.$\,\Psi^2$ and
 $\tilde{\Psi}^2$) have the same poles. Using this and the
asymptotic estimates at infinity we conclude that
$\Psi^1/\tilde{\Psi}^1$ and $\Psi^2/\tilde{\Psi}^2$ are
meromorphic functions on $C_h$ which have one pole (at
$\tilde{\Psi}^{i} = 0$). Moreover
$$
     \Psi_1(t,\infty^-) \big/
	   \tilde{\Psi}_1(t,\infty^-) = 1, \qquad
     \Psi_2(t,\infty^-) \big/
	   \tilde{\Psi}_2(t,\infty^-) = 1
$$
and hence $\Psi = \tilde{\Psi}$. At last the reader may check that
the functions (\ref{22a}) and (\ref{22b}) have the analyticity
properties from Proposition \ref{Ap2.pr} and hence they coincide
with the Baker--Akhiezer function defined in Proposition
\ref{Ap1.pr}. {\hfill $\Box $}


\subsection{Solutions of the Lagrange top}

      Let  $z=(z_1,z_2) \in J(C_h;\infty^\pm)$. It is easy to check
that the functions
$$
   \frac{\theta_{11}(z_1\pm \tau _2)}
	{\theta_{11}(z_1)} \, e^{\mp z_2}
$$
live on $J(C_h;\infty^\pm)$. We shall see that they give solutions
of the Lagrange top.  By (\ref{abel1}) we compute that
$\frac{d}{dt} z = {\rm constant}$, where
\vmmm
\ben
       \frac{dz}{dt}
     = \left(\begin{array}{c} V_1 \\ V_2 \end{array}\right)
     =2 \pi i \left(
	   \begin{array}{ll}
	       \int_{A_1}\frac{d\lambda }{\mu } 	 &
	       \int_{A_2}\frac{d\lambda }{\mu }
\\[8pt]
	       \int_{A_1} \frac{\lambda d\lambda }{\mu } &
	       \int_{A_2} \frac{\lambda d\lambda }{\mu }
	   \end{array}
	       \right)^{-1}
	       \left(
	   \begin{array}{c}
		 -i	      \\[5pt]
		-a i
	   \end{array}
	       \right) ,\;
\\[8pt]
       \int_{A_2} \frac{d\lambda }{\mu } = 0 \ , \qquad
       \int_{A_2} \frac{\lambda d\lambda }{\mu }
     = -2 \pi i
\een
so
$$
	 \left( \begin{array}{c} V_1 \\ V_2 \end{array} \right)
      =  \left(
		 \int_{A_1} \frac{ d\lambda }{\mu }
	 \right)^{-1}
	 \left( \begin{array}{c}
		    2\pi				 \\[5pt]
		    - i \int_{A_1} \frac{\lambda d\lambda}{\mu }
		    + a i \int_{A_1} \frac{ d\lambda }{\mu }
		\end{array}
	 \right)  \ ,
     \qquad
     a = -m \Omega_3 \;.
$$

\begin{th1}
  \label{theta.th}
The following equations hold
\be
  \label{om1}
	\overline{\epsi} \Omega_1(t) + \epsi \Omega_2(t)
      = {\rm const}_3	  \,
	  \frac{\theta_{11} (z_1 - \tau _2)}
	       {\theta_{11} (z_1)} \,  e^{- z_2} \ \
\\[7pt] \label{om2}
	\epsi \Omega_1(t) + \overline{\epsi} \Omega_2(t)
      = {\rm const}_4	   \,
	  \frac{\theta_{11} (z_1 + \tau _2)}
	       {\theta_{11} (z_1)} \, e^{+ z_2}  \ ,
\ee

\vspace{-0.7em}

\noindent
where

\vspace{-1.5em}

\begin{equation}
       z_2  =  t V_2,				       \quad
       z_1  =  t V_1 + {\cal A} (\infty^+ +\infty^-
		     - P_1 - P_2),		       \quad
     \tau_2 =  {\cal A} (\infty^+ - \infty^-)
	    = \int_{B_1} \omega_2
\end{equation}

\vspace{-0.7em}

\noindent
and

\vspace{-1.5em}

\ben
      {\rm const}_3
  &=&  \frac{2i\, V_1\, \theta_{11}^{'}(0)}
	    {\theta_{11}\big({\cal A}(\infty^--\infty^+)\big)}\cdot
       \frac{\theta_{11}\big({\cal A} (\infty^+-P_1)\big)}
	    {\theta_{11}\big({\cal A}(\infty^--P_1)\big)} \cdot
       \frac{\theta_{11}\big({\cal A} (\infty^+-P_2)\big)}
	    {\theta_{11}\big({\cal A}(\infty^--P_2)\big)}
\\[6pt]
      {\rm const}_4
  &=&  \frac{2i \, V_1 \, \theta_{11}^{'}(0)}
	    {\theta_{11}\bigl(
	      {\cal A} (\infty^+ -\infty^-)\bigr) }\cdot
       \frac{\theta_{11}\bigl(
	      {\cal A} (\infty^- -P_1)\bigr) }
	    {\theta_{11}\bigl(
	      {\cal A}(\infty^+  -P_1)\bigr) } \cdot
       \frac{\theta_{11}\bigl(
	      {\cal A} (\infty^- -P_2)\bigr) }
	    {\theta_{11}\bigl(
	      {\cal A}(\infty^+  -P_2)\bigr) }	 \ .
\een
\end{th1}

\noindent
Let us denote
\vspace{-1em}

\begin{align}
   \nonumber
      \omega_1
  & = \pm \bigl( \omega_1^0 + O(\lambda^{-1}) \bigr) \,
      d \bigl( \lambda^{-1} \bigr) , \
   &  P = (\lambda, \mu ) & \rightarrow \infty^{\pm}
\\[5pt] \nonumber
      \omega_2
  & = \pm \bigl( \omega_2^1 \lambda
	  + \omega_2^0 + O(\lambda^{-1})  \bigr)     \,
     d \bigl( \lambda^{-1} \bigr) ,   \
   &  P = (\lambda ,\mu ) & \rightarrow \infty^{\pm}
\end{align}
To prove Theorem \ref{theta.th} we shall need the following

\mm

\begin{lm}
  \label{const.lm}
  The above defined differentials are such that

    \vspace{-0.5cm}

\ben
 & & \omega_1^0 = - i \int_{B_1} \Omega = -i V_1 \,,\qquad
     \omega_2^0 = i \, (c^+ - c^-) \,,
  \\
 & & V_2 = -c^+ + c^- + i \Omega_3	\,,  \qquad
     {\cal A}(\infty^+ - \infty^-) = \int_{B_1} \omega _2 \ .
\een
\end{lm}

\noindent
{\bf {Proof.}}
The identity $\omega _1^0=-i\int_{B_1} \Omega $ is a
reciprocity law between the differential of first kind
$\omega _1$ and the differential of second kind $\Omega $
\cite{GH}. It is obtained by integrating $\pi (P) \omega _1$,
where  $\pi (P)=\int_{P_0}^P \Omega$, along the border of $C_h$
cut along its homology basis $A_1$, $B_1$. On the other hand
$$
	 \omega_1 = 2 \pi i
		      \left(
		      \int_{A_1} \frac{d\lambda }{\mu }
		      \right)^{-1}
		   \frac{d \lambda }{\mu }
$$
and hence
$$
	 \omega_1^0 = -2 \pi i
	 \left(
	 \int_{A_1} \frac{d\lambda }{\mu}
	 \right)^{-1}
       = - i V_1 \; .
$$
Similarly the identity $\omega_2^0 = i(c^+ - c^-)$ is a reciprocity
law between the differential of third kind $\omega _2$
and the differential of second kind $\Omega $, and
${\cal A}(\infty^+-\infty^-) = \int_{B_1} \omega _2 $ is a
reciprocity law between the differential of third kind
$\omega _2$ and the differential of first kind $\omega_1$.
At last as
$$
    \omega_2  \ = \ \frac{\int_{A_1} \frac{\lambda d \lambda}
					  {\mu }}{\int_{A_1}
		    \frac{d \lambda }{\mu}}
		    \frac{d \lambda }{\mu}
		  - \frac{\lambda d \lambda}{\mu}
$$
\vspace{-0.3mm}
then
$$
	  \omega_2^0 \
    = \ - \frac{\int_{A_1} \frac{\lambda d \lambda}{\mu}}
	       {\int_{A_1} \frac{d \lambda}{\mu}}
	- (1+m) \Omega_3 \
    = \ -i V_1 - \Omega_3
$$
and hence $V_2 = -c^+ + c^- + i \Omega_3$ . {\hfill $\Box$}

\mm
\noindent
{\bf {Proof of Theorem \ref{theta.th}.}}
According to (\ref{e15}), (\ref{e16})
$$
	 \overline{\epsi} \Omega_1(t) + \epsi \Omega_2(t)
      = -2 \lim_{P \rightarrow \infty^-}
	      \frac{\lambda \Psi^1(t,P)}{\Psi^2(t,P)}  \ \
$$
\vspace{-14mm} $$ $$
and
$$
	 \epsi \Omega_1(t) + \overline{\epsi} \Omega _2(t)
      = +2 \lim_{P \rightarrow \infty^+}
	      \frac{\lambda \Psi^2(t,P)}{\Psi^1(t,P)} \; .
$$
To compute the limit we use (\ref{22a}), (\ref{22b}) and
$$
       \lim_{P \rightarrow \infty^-}
	  \lambda (P) \, \theta_{11}\big({\cal A}(P-\infty^-)\big)
	= \theta_{11}^{'}(0) \, \frac{d}{d s}_{\big | s=0}
	  \int^s \omega_1
	= \omega_1^0 \, \theta_{11}^{'}(0)
$$
$$
       \lim_{P \rightarrow \infty^+}
	  \lambda (P) \, \theta_{11}\big({\cal A}(P - \infty^+)\big)
	= \theta_{11}^{'}(0) \, \frac{d}{ds}_{\big | s=0}
	  \int^s \omega_1
	= \omega_1^0 \, \theta_{11}^{'}(0)
$$
(see Lemma \ref{const.lm}). {\hfill $\Box$


\subsection{Effectivization}
  \label{effectivization}

     Let $\wp, \zeta , \sigma$ be the  Weierstrass functions
related to the elliptic curve $\Gamma$ defined by
\begin{equation}
  \label{we}
	 \eta^2  =  4 \xi^3 - g_2 \xi - g_3
\end{equation}
(we use the standard notations of \cite{Erdely}).

     Consider also the {\it real } elliptic curve $C$ with
affine equation
\begin{equation}
  \label{ew}
	  \mu^2 + \lambda^4    + a_1 \lambda^3 + a_2 \lambda ^2
		+ a_3 \lambda  + a_4		=  0
\end{equation}
and natural anti--holomorphic involution
$(\lambda,\mu) \rightarrow
\big(\overline{\lambda},\overline{\mu}\big)$,
and put
\begin{equation}
   \label{g2g3}
       g_2  =  a_4 + 3 \Big( \frac{a_2}{6} \Big)^4
		   - 4 \frac{a_1}{4} \frac{a_3}{4},  \quad
       g_3  =  {\rm det}
		    \left(
	    \begin{array}{ccc}
	       1	     & \tfrac{a_1}{4} & \tfrac{a_2}{6} \\
	      \tfrac{a_1}{4} & \tfrac{a_2}{6} & \tfrac{a_3}{4} \\
	      \tfrac{a_2}{6} & \tfrac{a_3}{4} & a_4
	    \end{array}
		    \right) \; .
\end{equation}
It is well known that the curves $C$ and $\Gamma$ are isomorphic
over $\CC$ and that under this isomorphism
\vmmm
\begin{equation}
  \label{hd}
      \frac{d\lambda}{\mu}  =  \frac{d\xi} { \eta } \quad .
\end{equation}
Following Weil \cite{Weil} we call $\Gamma$ Jacobian $J(C)$ of
the elliptic curve $C$ and we write $J(C)=\Gamma$. Note that
$J(C)$ and $\Gamma$ are real isomorphic and that $J(C)$ and $C$
are not real isomorphic.

    Further we make the substitution (\ref{eah}) and $C$ becomes
the spectral curve $\tilde{C}_h$ of Adler and van Moerbeke
$\{\mu^2+f(\lambda)=0 \} $ where
$$
       f(\lambda)  =  \lambda^4 + 2(1+m)h_4 \lambda^3
		     + \big(2h_3 + m(m+1)h_4^2\big) \lambda^2
		     - 2 h_2 \lambda +1
$$
and $\Gamma$ becomes the Lagrange curve $\Gamma_h$. Recall
that, as we explained at the end of section \ref{algebraic},
the curve $C_h$ with an equation $\{\mu^2=f(\lambda) \}$ and
antiholomorphic involution
$(\lambda, \mu) \rightarrow (\overline{\lambda},-\overline{ \mu})$,
is isomorphic over $\RR$ to
$\tilde{C}_h$, so we write $C_h=\tilde{C}_h$.
The Jacobian curve $J(C_h) = \Gamma_h$ was computed by Lagrange
\cite{Lagrange}, while	$C_h$ appeared first in \cite{Adler,Ratiu}
as a spectral curve of a Lax pair associated to the Lagrange top.

      Recall that  $\sigma (z)$ is an entire function in $z$
related to $\zeta (z), \wp(z)$ and the already defined function
$\theta_{11}(z|\tau_1)$ on $C_h$ as follows
\vmm
\be
   \zeta '(z)                      =  -\wp(z) \ , \qquad
   \frac{\sigma '(z)}{\sigma (z)}  =  \zeta (z)\ ,  \qquad
				  '= \frac{d}{dz}  \qquad \
\nonumber
\\[5pt]
  \label{sth}
     \sigma (z) = \frac{\theta_{11}(zU)}{U\theta_{11}'(0)}
     \exp \left\{ \frac{z^2 U^2 \theta_{11}'''(0)}
		       {6 \theta_{11}'(0)} \right \}
   =  z - \frac{g_2 z^5}{240} + \cdots
\ee
where $U$ is a constant depending on $g_2$ and $g_3$.
We shall also use the ``addition formula''
$$
       \frac{\sigma (u+v) \, \sigma (u-v)}
	    {\sigma ^2(u) \, \sigma ^2(v)}
    \ = \ \wp (v) - \wp(u) \ .
$$
To state our result let us introduce the notations
\begin{align}
    2 x_1 &= \epsi \Omega_1 + \overline{\epsi}\Omega_2&
    2 x_2 &= \overline{\epsi}\Omega_1 + \epsi \Omega_2&
	   \epsi^2= \sqrt{-1}
\nonumber \\
    2 y_1 &= \epsi^3 \Gamma_1 + \epsi	\Gamma_2      &
    2 y_2 &= \epsi \Gamma_1   + \epsi^3 \Gamma_2      &
	    i^2 = -1  \ \
  \label{xy} \\
    \rho_1&= - i \, m \, \Omega_3		      &
    \rho_2&= - i \,  \Omega_3	       \ .	      &
  \nonumber
\end{align}
The system (\ref{Ltop}) is equivalent to
\begin{align}		    \nonumber
     \bp{x}_1	   &\, =\, + \rho _1 x_1 - y_1		  &
     \bp{y}_1	   &\, =\,    - \rho_2 y_1 + x_1 \Gamma_3 &
 \\			     \label{top}
     \bp{x}_2	   &\, =\,   - \rho_1 x_2 + y_2 	  &
     \bp{y}_2	   &\, =\,    + \rho_2 y_2 - x_2 \Gamma_3 &
     \\ 		   \nonumber
 \rho_1\,,\,\rho_2 &\, =\,    {\rm constants}		  &
    \bp{\Gamma}_3  &\, =\      2 x_1 y_2  -  2 x_2 y_1	  &
\end{align}
with first integrals
$ I_0 = 4 x_1 x_2 - 2 \Gamma_3, 	       \,
  I_1 = 4 x_1 y_2 + 4 x_2 y_1 -2 (\rho _1+\rho _2)\Gamma _3 $
\, and \,
$ I_2 = \Gamma _3^2 - 4 y_1 y_2 .$

\begin{th1}
  \label{sigma.th}
The general solution of the Lagrange top {\rm (\ref{Ltop})}
can be written in the following form
\begin{align}
      x_1(t) &= - \, \frac{\sigma (t-k-l)}
			   {\sigma (t) \,\sigma (k+l)} \,
			   e^{at+b}  &
      x_2(t) &= - \, \frac{\sigma (t+k+l)}
			   {\sigma (t) \,\sigma (k+l)} \,
			   e^{-at-b} &
\nonumber
\\[7pt]
      y_1(t) &= \ \, \frac{\sigma (t-k)\,\sigma (t-l)}
			{\sigma^2(t)\,\sigma (k)\,\sigma (l)}\,
		       e^{at+b}  &
      y_2(t) &= \ \, \frac{\sigma (t+k)\,\sigma (t+l)}
			{\sigma ^2(t)\,\sigma (k)\,\sigma (l)}\,
		       e^{-at-b} &
  \nonumber
\end{align}
$$
      \  \Gamma_3(t)	   \,
    = \, \frac{\sigma (t+k) \, \sigma (t-k)}
	      {\sigma ^2(k) \, \sigma ^2(t)}
       + \frac{\sigma (t+l) \, \sigma (t-l)}
	      {\sigma ^2(l) \, \sigma ^2(t)}  \,
    = \, -2 \wp(t) + \wp(l) + \wp(k) \qquad \qquad
$$
\vspace{-5mm}
\ben
      \rho_1 =	a - \zeta (l) - \zeta (k)  \hspace{32mm}
      \rho_2 = -a - \zeta (k) - \zeta (l) + 2 \zeta (k+l) \ ,
      \hspace{-9mm}
\een
where $g_2, g_3, a, b, k, l$ are arbitrary constants subject
to the relation $ g_2^3 - 27 g_3^2 \neq 0 $.
\end{th1}
{\bf Remark}. The non--general solutions of the Lagrange top are
obtained from the above formulae by taking the limit
$\, g_2^3 - 27 g_3^2 \rightarrow 0 $. The formulae for the
position of the body in space, and in particular for
$\Gamma_3(t), y_1(t), y_2(t)$, are due to  Jacobi \cite{Jacobi2}.
The expressions for $x_1(t), x_2(t)$ were first deduced by Klein
and Sommerfeld \cite[p.436]{Klein}. Note however that in
\cite{Klein} the constant $a$, and hence the invariant level set
on which the solution lives, is not arbitrary.

\mm
\noindent
{\bf {Proof.}}
To make the solutions of the Lagrange top effective we use the following
dimension four Lie group of transformations preserving the
system (\ref{top})
\vmm
\begin{align}
      x_1 &\rightarrow	U x_1 e^{at+b}, 	       &
      x_2 &\rightarrow	Ux_2 e^{-at-b}, 	       &
      t \ &\rightarrow	\frac{t}{U} + T    \nonumber   &\\
      y_1 &\rightarrow	U^2 y_1 \, e^{at+b},	       &
      y_2 &\rightarrow	U^2 y_2 \, e^{-at-b},	       &
  \Gamma_3&\rightarrow	U^2 \, \Gamma_3    \label{Lie} &\\[6pt]
    \rho_1&\rightarrow	U \rho_1 + a,		       &
    \rho_2&\rightarrow	U \rho_2 - a	   \nonumber   & &
\end{align}
where $U \neq 0, T,a,b$ are constants.

     The group (\ref{Lie}) transforms $x_1$
from (\ref{om1}) \big(see also (\ref{xy}), (\ref{sth})\big),
where $z_1 = tU - TU$, $z_1 - \tau_2 = (t-k-l)U$ as follows
$$
    x_1(t) = {\rm const}			     \,
	     \frac{\theta_{11}(z_1-\tau_2)}
		  {\theta_{11}(z_1)}
	   = - \,\frac{\sigma (t-k-l)}
		       {\sigma (t) \, \sigma (k+l)}  \,
	      e^{at+b} \, .
$$
(we used that
$$
	\frac{\theta_{11}(z_1-\tau_2) \, \sigma (t)}
	     {\theta_{11}(z_1)	      \, \sigma (t-k-l)}
$$
is a constant). The variable $x_2$ is computed in the same way.

     If we define  the constant $k$ by the condition
$y_1(t-k)=0$, then the first equation of (\ref{top}) gives
$$
	  \frac{y_1(t)}{x_1(t)}
       =  \rho_1 -\frac{x'_1(t)}{x_1(t)}
       =  \frac{\sigma(t-k) \, h(t)}
	       {\sigma(t)   \, \sigma(t-k-l)}
$$
where $h(t)$ is a meromorphic function on $\CC$, such that
$y_1(t)/x_1(t)$ is singe valued with poles at $t=0$ and
$t=k+l$, and residues $(-1)$ and $(+1)$ respectively.
These three conditions define $h(t)$ uniquely:
$$
      h(t) = \frac{\sigma(t-l) \, \sigma(k+l)}
		  {\sigma(k)   \, \sigma(l)} \ ,
$$
which implies the formula for $y_1(t)$. The expression for $y_2(t)$
is obtained in the same way.

     To deduce an expression for $\Gamma_3(t)$ we use that
$$
       \Gamma _3(t)
     =	2 x_1 x_2 - \tfrac{1}{2} I_0
     = -2 \wp(t) + 2 \wp(k+l) - \tfrac{1}{2} I_0 \ .
$$
The value of $I_0$ is easily computed by  making use  the third
equation of (\ref{top}) and the formulae deduced for $x_1, y_1$.
By substituting $t=k$ we obtain
$$
	  \Gamma_3(k)
       =  \frac{\sigma (k-l) \, \sigma (k+l)}
	       {\sigma ^2(k) \, \sigma^2 (l)}
       =  \wp(l) - \wp(k)
$$
and in a similar way $\Gamma_3(l) = \wp(k)-\wp(l)$. We conclude
that
$$
       \Gamma _3(t) = -2 \wp(t) + \wp(l) + \wp(k) \ .
$$

       At last, to compute $\rho_1, \rho_2$ we shall use once
again (\ref{top}). As $y_1(k) = 0$ then
\ben
      \rho_1 &=& \frac{\bp{x}_1(k)}{x_1(k)}
	      =  \frac{d}{dt} \ln x_1(t)_{\big|{t=k}}
\\	     &=& \frac{d}{dt} \ln \sigma (t-k-l)_{\big|{t=k}}
		-\frac{d}{dt} \ln \sigma (t)_{\big|{t=k}} + a
\\	     &=& a-\zeta (l)- \zeta (k) \; .
\een
In a quite similar way we obtain
$$
	      \rho _2
      =     - \frac{d}{dt} \ln y_1(t)_{\big|{t=k+l}}
      = - a - \zeta (k)- \zeta (l) + 2 \zeta (k+l) \ .
$$
Theorem \ref{sigma.th} is proved.

\mm  \noindent
{\bf Remark.} If we impose the condition
\ben
	\Gamma_1^2 + \Gamma_2^2 + \Gamma_3^2
     =	\Gamma_3^2 - 4 y_1 y_2
     =	1    \ ,
\een
\vspace{-15mm} $$ $$
then
\vm
\ben
     \left(  \frac{\sigma (t+k) \, \sigma (t-k)}
		  {\sigma ^2(k) \, \sigma ^2(t)}
	  +  \frac{\sigma (t+l) \, \sigma (t-l)}
		  {\sigma ^2(l) \, \sigma ^2(t)}
     \right)^2
	  -  \frac{\sigma (t-k) \, \sigma (t-l)}
		  {\sigma ^2(t) \, \sigma (k) \, \sigma (l)}
	     \frac{\sigma (t+k) \, \sigma (t+l)}
		  {\sigma ^2(t) \, \sigma (k) \, \sigma (l)}
\\[4pt]
   = \left ( \frac{\sigma (t+k) \, \sigma (t-k)}
		  {\sigma ^2(k) \, \sigma ^2(t)}
	   - \frac{\sigma (t+l) \, \sigma (t-l)}
		  {\sigma ^2(l) \, \sigma ^2(t)}
     \right ) ^2
   = \big(\wp(k) - \wp(l)\big)^2
   = 1
\een
and hence $\wp(k) - \wp(l) = \pm 1$.


\section{Real structures}
\label{real}

      Recall that a {\it real algebraic variety} is a pair
$(X,S)$ where $X$ is a complex algebraic variety and
$S: X\rightarrow X$ is an anti--holomorphic involution on it.
The set of fixed points of $S$ is the {\it real part} of
$(X,S)$. $S$ acts on the group of divisors ${\rm Div(X)}$ :
if $D\in {\rm Div(X)}$ is defined locally by analytic
functions $f_\alpha $, then $S(D)$ is defined by the
analytic functions $\overline{f_\alpha \circ S}$. Thus it
is natural to define an involution $S^*$ on the sheaf of
analytic functions ${\cal O}_X$
$$
	S^* : \, \Gamma \,\big( S(U),{\cal O}_X \big)
		 \rightarrow
		 \Gamma \big( U,{\cal O}_X \big) \,
	    : \, f \mapsto \overline{f\circ S} \ .
$$
This induces also an involution on the group of one--forms and
one--cycles. If $\omega \in H^0(X,\Omega ^1)$,
$c \in H_1(X,\ZZ)$  then
$\int_c S^*\omega  = \overline{\int_{S(c)}\omega }$.
A form $\omega $ is $S$--real if and only if $S^* \omega = \omega $
and one may always choose a basis of $S$--real forms.
In the case when $X=C_h$ is the spectral curve of the Lagrange top,
the action of $S$ on ${\rm Div(X)}$ induces an involution on
$J(C_h;\infty^\pm)$. This, however, does not suffice to
determine the real structure of the invariant manifold
$T_h \sim J(C_h;\infty^\pm) \setminus \phi^{-1}(p)$
(Theorem \ref{th.Lagrange}), as it will also depend on the point
$p\in J(C_h)$. Recall that the symmetric product
$S^2 \breve{C}_h$ is bi--rational to $T_h$. Thus the generalized
Jacobian and the invariant manifold $T_h$ are identified by
the Abel map
\begin{equation}
  \label{eqs2}
	    {\cal A} : \, S^2 \breve{C}_h     \rightarrow
  J(C_h; \infty^\pm) \ : \, P_1 + P_2	      \mapsto
    \int_{W_1+W_2}^{P_1+P_2} \omega \ ,       \qquad
		 \omega = (\omega_1 , \omega_2) \ .
\end{equation}
This induces an involution on $J(C_h;\infty^\pm)$,
$z \rightarrow S(z)$, where
$$
	 z = \int_{W_1 + W_2}^{P_1 + P_2}    \omega \ , \qquad
      S(z) = \int_{W_1 + W_2}^{S(P_1 + P_2)} \omega \ .
$$
Of course this depends on the fixed points $W_1, W_2 \in
J(C_h;\infty^\pm)$. Let $\omega _1, \omega _2$ be $S$--real . Then
$$
      S(z) = \int_{W_1 + W_2}^{S(W_1 + W_2)}	\omega
	   + \int_{S(W_1 + W_2)}^{S(P_1 + P_2)} \omega
	\, = \, \int_{W_1 + W_2}^{S(W_1 + W_2)}    \omega
	   + \overline{\int_{W_1 + W_2}^{P_1 + P_2} \omega}
	\, = \, S(0) + \overline{z} \ .
$$
If $S$ has a fixed point on  $J(C_h;\infty^\pm)$ (this does not
depend on $W_1$, $W_2$) then one may always choose it for
origin, and hence $S(z)=\overline{z}$ becomes a group homomorphism.

     Denote by $S$ the anti--holomorphic involution on the spectral
curve $C_h$ defined by	
$ S(\lambda, \mu) = (\overline{\lambda}, - \overline{\mu})$. This
involution comes from the real Lax pair of Adler and van Moerbeke
defined in section \ref{algebraic}.  We shall also suppose that
the real polynomial $f(\lambda )$ has distinct roots. S
induces an involution on the usual Jacobian $J(C_h)$ which we
denote by $S$ too, and an involution on the generalized
Jacobian $J(C_h;\infty^\pm)$ which we denote by $S^+$. If we
use (\ref{eqs2}), then in terms of the Jacobi polynomials $U,V,W$,
it is given by
$$
    S^+ : \, \big(U,V,W \big)
	     \mapsto
    \big( \, \overline{U}, -\overline{V}, \overline{W} \, \big) .
$$
There is another natural anti--holomorphic involution on $T_h$
given by the usual complex conjugation
$$
      \big(   \Omega_i , \Gamma_i \big)
	      \mapsto
      \big( \,\overline{\Omega}_i , \overline{\Gamma}_i \, \big)
$$
which we denote by $S^-$. In terms of the Jacobi polynomials
(\ref{jp}) it is
$$
      S^- : \, \big( U,V,W \big)
	    \mapsto	   \big( \,
	    \overline{W}, \overline{V}, \overline{U} \,
			   \big) \, .
$$

\begin{pr}
  \label{real.pr}
The holomorphic involution $S^+ \circ S^- = S^- \circ S^+$ on
$J(C_h;\infty^\pm)$  is a translation on the half--period
$\, \tfrac{1}{2}\Lambda_2$, where $\phi \big( \tfrac{1}{2}
\Lambda_2 \big) = 0 \in J(C_h)$
{\rm \big(}see {\rm (\ref{extension}),(\ref{extension2})\big)}.
\end{pr}
The proof of the above Proposition will be given later in this
section. If $\phi $ is the projection homomorphism defined in
(\ref{extension}), then it implies
$$
     \phi \circ S^+  =	\phi \circ S^-	=  S\circ\phi \; .
$$
In other words the anti--holomorphic involutions $S^+$ and $S^-$
``look like'' in the same way on the usual Jacobian $J(C_h)$
and differ in a half--period in the ``vertical'' direction
with respect to $\phi $ on the generalized Jacobian
$J(C_h;\infty^\pm)$.

     An important feature of $S^+$ is that the $S^+$--real part
of the invariant level set $T_h$ is preserved by the flow of
(\ref{Ltop}). Indeed, changing the variables as
\begin{align}			       \nonumber
	\Omega_1 &\rightarrow i \Omega_1&
	\Omega_2 &\rightarrow i \Omega_2&
	\Omega_3 &\rightarrow	\Omega_3&
\\				       \nonumber
	\Gamma_1 &\rightarrow i \Gamma_1&
	\Gamma_2 &\rightarrow i \Gamma_2&
	\Gamma_3 &\rightarrow	\Gamma_3&
\end{align}
we obtain a new system
\begin{align}
     \bp{\Omega }_1  & =  -m\, \Omega_2\Omega_3 - \Gamma_2 &
     \bp{\Gamma}_1   & =      \Gamma_2 \Omega_3
			    - \Gamma_3 \Omega_2 	   &
\nonumber
\\ \label{Ltop*}
     \bp{\Omega }_2  & = \ \ m\,\Omega_3\Omega_1 + \Gamma_1&
     \bp{\Gamma}_2   & =      \Gamma_3 \Omega_1
			    - \Gamma_1 \Omega_3 	   &
\\
     \bp{\Omega }_3  & =  \quad  0			   &
     \bp{\Gamma}_3   & =     \Gamma_2 \Omega_1
			    - \Gamma_1 \Omega_2 	   &
\nonumber
\end{align}
with first integrals
\begin{align}
      H_1 & = - \Gamma_1^2 - \Gamma_2^2 + \Gamma_3^2	   &
      H_2 & = - \Omega_1 \Gamma_1 - \Omega _2\Gamma_2
	      + (1+m) \Omega_3 \Gamma_3 &
\nonumber
\\
      H_3 & = \tfrac{1}{2} \, \big( -\Omega_1^2 -\Omega_2^2
		+ (1+m) \Omega_3^2 \, \big)	-\Gamma_3  &
      H_4 & = \Omega _3 \  .				   &
\nonumber
\end{align}
The anti--holomorphic involution $S^+$ in these  coordinates
is given again by the complex conjugation.

\begin{th1}
\label{real.th}
In each of the three connected subdomains of the complement
to the discriminant locus of $f(\lambda )$ the topological
type of the real part of the algebraic varieties
\\ $\big( J(C_h;\infty^\pm),S^\pm \big)$ and $(T_h,S^\pm)$
is one and the same and it is given in the following table

\bigskip

\begin{tabular}{|c|c|c|c|}
\hline
   roots of $f(\lambda )$
&  no real roots
&  two real roots
&  four real roots
\\[2pt]
\hline	\\[-25pt] & & & \\
real part of  $ \big( J(C_h;\infty^\pm),S^+ \big) $
&  $T^2$
&  $T^2$
&  $T^2 \times (\ZZ /2)$
\\[2pt]
\hline	\\[-25pt] & & & \\
   real part of  $\big(J(C_h;\infty^\pm),S^-\big)$
&  $T^2$
&  $\emptyset$
&  $\emptyset$
\\[2pt]
\hline	\\[-25pt] & & & \\
   real part of  $\big(T_h,S^+\big)$
&  $S^1 \times \RR$
&  $S^1 \times \RR$
&  $T^2 \cup (S^1 \times \RR) $
\\[2pt]
\hline	\\[-25pt] & & & \\
   real part of  $(T_h,S^-)$
&  $T^2$
&  $\emptyset$
&  $\emptyset$
\\[2pt]
\hline
\end{tabular}
\\
\newline
where $T^2 = S^1 \times S^1$ .
\end{th1}

\noindent
{\bf Remark.} It is easy to check that when the real
invariant level set $T_h^\RR$ of the Lagrange top is non--empty,
then the polynomial $f(\lambda)$ has no real roots.
If we do not use the generalized Jacobian
$J(C_h;\infty^\pm)$, then it might be difficult to
understand the relation between $T_h^{\RR}$ (which has one
connected component), $C_h^\RR$ (which is empty) and
$J(C_h)^{\RR}$ (which has two connected components)
\cite{Audin1}, \cite[p.37]{Audin2}).

\mm
\noindent
{\bf {Proof of Proposition \ref{real.pr}.}}
We have $S^+ \circ S^- : (U,V,W) \mapsto (W,-V,U)$. The
involution $(U,V,W) \mapsto (U,-V,W) $ is obviously induced
by the elliptic involution $i:(\lambda ,\mu ) \mapsto
(\lambda ,-\mu )$ on $C_h$ so it is  a reflexion. This
means that if a fixed point of $i$ is taken for origin in
$J(C_h ; \infty^{\pm})$ then $i = - {\rm identity}$.
It remains to prove
that $j: (U,V,W) \mapsto (W,V,U)$ is a reflexion too. The
involution $j$ has the following simple geometrical interpretation.
Let $P_1, P_2$ be two generic points on the $(\lambda ,\mu )$
plane and lying on the affine curve
$\breve{C}_h = \{\mu ^2 = f(\lambda ) \}$.
If $\{\mu =V(\lambda ) \}$ is the straight line through $P_1$
and $P_2$ then it intersects $C_h$ in four points
$P_1,P_2,P_3,P_4$
and then $j(P_1+P_2)=P_3+P_4$.
Indeed, if the zero divisor of the Jacobi polynomial
$U(\lambda )$ on $C_h$ is $P_1 + P_2 + i(P_1) + i(P_2)$,
then by (\ref{jp1}) the zero divisor of $W(\lambda )$ is
$P_3 + P_4 + i(P_3) + i(P_4)$
and the involution $P_1+P_2 \mapsto P_3+P_4$ amounts to
exchanging the roots of $U(\lambda )$ and $V(\lambda )$.

     Let $W_i$, $i=1,..,4$ be the Weierstrass points on
$C_h$. Then
\vm
$$
       \left(
	 \frac{\mu -V(\lambda )}{\mu } \right)
     \	= \ \sum_{i=1}^4 P_i - \sum_{i=1}^4 W_i \ , \qquad
	 \frac{\mu - V(\lambda) }{\mu }
     \	\stackrel{\infty}{\sim}  \  1
$$
\vspace{-8mm} $$ $$
and hence on
$ J (C_h ; \infty^{\pm}) \sim {\rm Div}^0(\breve{C}_h)
 / \stackrel{m}{\sim} \ $
holds $P_1 + P_2 = - P_3 - P_4 + {\rm constant\,}.$
This implies that $j$ is a
reflexion. Thus we proved that $S^+ \circ S^-$ is a
translation $\big( S^+ \circ S^- \big) (z) = z + a$.
At last $a$ is easily computed. We have
$i(W_k) = W_k$, $j(W_1 + W_2) = W_3 + W_4$ and hence
$a \stackrel{m}{\sim} W_1+W_2-W_3-W_4$. Further if
$\lambda_1, \lambda_2$	are zeros of  $f(\lambda )$,
then $(g) = W_1 + W_2 - W_3 - W_4$, where
$g(\lambda ) = (\lambda - \lambda_1)
 (\lambda - \lambda_2)/\mu $.
Moreover $g(\infty^\pm)=\pm1$, $g^2(\infty^\pm)= 1$
and hence
\vm
$$
    W_1 + W_2 - W_3 - W_4		  \sim 0 \ , \quad
    W_1 + W_2 - W_3 - W_4  \stackrel{m}{\not \sim}0, \quad
2\,(W_1 + W_2 - W_3 - W_4) \stackrel{m}{\sim}	   0 \ .
$$
This shows that $a$ is a half--period and
$\phi(a)=0 \in J(C_h)$.
{\hfill $\Box $}

\mm
\noindent
{\bf {Proof of Theorem \ref{real.th}.}}
The proof will consist of two  steps. First we determine the
action of $S^\pm$ on $H_1(\breve{C}_h,\ZZ)$ and hence on
the period lattice $\Lambda $. From that we deduce the
first two lines of the table. Second, we determine the action
of  $S^\pm : D_\infty \mapsto D_\infty$ on the infinity
divisor $D_\infty=\phi^{-1}(p)= \CC^2/ \Lambda _2 \sim C^*$ and
then we use that
\begin{center}
	real part of $\big(T_h,S^\pm\big)$
  \ = \ real part of
	$\big( J(C_h;\infty^{\pm}),S^{\pm} \big)$
	$\, - \,$  real part of $D_\infty$ .
\end{center}

     It is easier to determine the action of $S^+$ on $\Lambda $.
Indeed, $S^+$ is induced by an anti--holomorphic involution on
$C_h$, $S^+: \big( \lambda ,\mu \big) \mapsto
\big( \, \overline{\lambda}, -\overline{\mu} \, \big)$.
Note that $S^+$ has always fixed points
on $J(C_h;\infty^\pm)$ : if $W_1,W_2$ are two Weierstrass points on
$C_h$ such that either $W_1 = \overline{W}_2$,
or $W_1$ and $W_2$ are
$S^+$--real, then $S^+ (W_1 + W_2) = W_1 + W_2$. On the other hand
$S^-$ has fixed points only if $f(\lambda )$ has no real
roots. Indeed, in this last case let $W_i$, $i=1,..,4$, be the
Weierstrass points of $C_h$  where $W_1 = \overline{W}_2$,
$W_3 = \overline{W}_4$. Then $j(W_1 + W_3) = W_2 + W_4$
(see the proof of Proposition \ref{real.pr}) and hence
$S^- (W_1 + W_3) = W_1 + W_3$. On the other hand if
$U = \overline{W}$ and $V = \overline{V}$, then
\ben
	 V^2 (\lambda )    +   U(\lambda ) W(\lambda )
     = | V   (\lambda )|^2 + | U(\lambda) |^2
     = f (\lambda ) > 0     \  \  \qquad
	 \forall \; \lambda  \in \RR	 \ ,
\een
and hence $f(\lambda )$ has no real roots.

     Suppose first that $f(\lambda )$ has no real roots and
let us choose a basis $A_1$, $B_1$, $A_2$ of
$H_1(\breve{C_h}, \ZZ)$ as it is shown on figure 2
and figure 3.
  
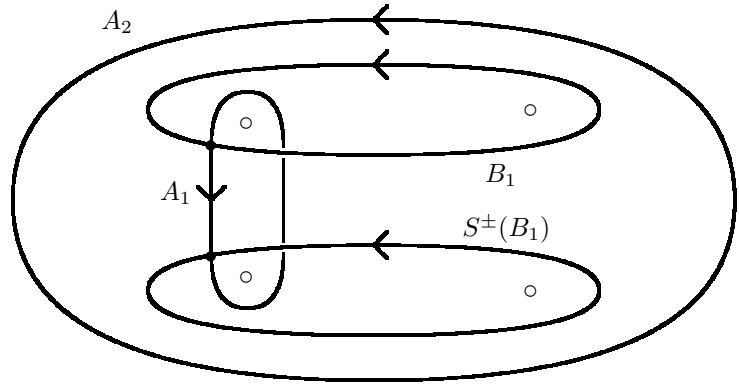
\vspace{1.5cm}
\begin{center}
\begin{picture}(60,80)(100,-40)
\setlength{\unitlength}{.6mm}
\linethickness{1pt}
\put(20,38){$A_2$}
\qbezier(80,40)(82,42)(83,43)
\qbezier(80,40)(82,38)(83,37)
\qbezier(80,30)(82,32)(83,33)
\qbezier(80,30)(82,28)(83,27)
\qbezier(80,-10)(82,-8)(83,-7)
\qbezier(80,-10)(82,-12)(83,-13)
    \qbezier(0,0)(0,40)(80,40)
    \qbezier(80,40)(160,40)(160,0)
    \qbezier(0,0)(0,-40)(80,-40)
    \qbezier(80,-40)(160,-40)(160,0)
\put(105,4){$B_1$}
    \qbezier(30,20)(30,30)(80,30)
    \qbezier(80,30)(130,30)(130,20)
    \qbezier(130,20)(130,10)(80,10)
    \qbezier(80,10)(30,10)(30,20)
\put(100,-8){$S^{\pm}(B_1)$}
    \qbezier(30,-20)(30,-30)(80,-30)
    \qbezier(80,-30)(130,-30)(130,-20)
    \qbezier(130,-20)(130,-10)(80,-10)
    \qbezier(80,-10)(30,-10)(30,-20)
\put(33,0){$A_1$}
     \qbezier(60,9)(60,0)(60,-9)
     \qbezier(44,12)(44,0)(44,-12)
     \qbezier(44,12)(44,24)(52,24)
     \qbezier(60,12)(60,24)(52,24)
     \qbezier(44,-12)(44,-24)(52,-24)
     \qbezier(60,-12)(60,-24)(52,-24)
   \qbezier(44,0)(46,2)(47,3)
   \qbezier(44,0)(42,2)(41,3)
\put(42.7,10.75){$\bullet$}
\put(42.7,-14){$\bullet$}
\put(52,17){\circle{2.2}}
\put(52,-17){\circle{2.2}}
\put(115,20){\circle{2.2}}
\put(115,-20){\circle{2.2}}
\end{picture}
\end{center}
\vspace{1cm}
\begin{center}
	    Fig.3: Projection of the cycles
	    $ A_1 , B_1 , A_2 , S^\pm(B_1) $ on the
	    $ \lambda$--plane
\end{center}

\vspace{.5cm}
Then $S^+ (A_1) = A_1$, $S^+ (A_2) = A_2$ and it is easily
seen that $S^+(B_1)+B_1$ is homologous to $ A_2$ on
$H_1(\breve{C_h},\ZZ)$. Thus in the basis $A_1,A_2, B_1$ the
matrix of the involution
$S^+: H_1(\breve{C_h},\ZZ)  \rightarrow H_1(\breve{C_h},\ZZ) $
takes the form
$$
      \left(
	    \begin{array}{lcr}
	       1     &	\  0	&     0        \\
	       0     &	\  1	&     1        \\
	       0     &	\  0	&    -1
	    \end{array}
      \right) \; .
$$
From this and the fact that $\big(J(C_h;\infty^\pm),S^+\big)$
is not empty we conclude that the real part of
$\big(J(C_h;\infty^\pm),S^+\big)$
is a torus with generators the periods $\int_{B_1}\omega\,$ and
$\,\int_{A_2}\omega $. On the other hand the real part of
$\big(J(C_h;\infty^\pm),S^-\big)$ is not empty too and
$S^+\circ S^-$ is a translation.  We conclude that the real part of
$\big(J(C_h;\infty^\pm),S^-\big)$ is just a translation of the real
part of $\big(J(C_h;\infty^\pm),S^+\big)$ and in particular it is
generated by the same periods.

       In a similar way we find the real part of
$\big(J(C_h ; \infty^{\pm}), S^+\big)$ in the remaining cases.
Note that in an appropriate $\ZZ$ basis of $H_1(\breve{C_h},\ZZ)$
the matrix of the involution
$S^\pm : H_1(\breve{C_h},\ZZ) \rightarrow H_1(\breve{C_h},\ZZ)$
takes the same form if $f(\lambda )$ has two real roots,
and it is of the form
$$
    \left(
	  \begin{array}{lcr}
	     1	  &   \ 0    &	 0     \\
	     0	  &   \ 1    &	 0     \\
	     0	  &   \ 0    &	-1
	  \end{array}
    \right) \;
$$
if $f(\lambda )$ has four real roots. This implies the first
two lines of the table.

      Let us determine now the real part of $(D_\infty ,S^\pm)$.
As $D_\infty \! = \CC^* / \Lambda_2$ then we have to compute
$S^\pm (\Lambda _2)$. Note that, as the real invariant
manifold $T_h$ is  compact, then $(D_\infty ,S^-)$ is always
empty. On the other hand $(D_\infty ,S^+)$ is never empty.
Indeed, if $S^+(\lambda ,\mu ) = (\overline{\lambda },
- \overline{\mu })$ then for $Q\in C_h$ the point
$Q + S^+ (Q)$  is  $S^+$--real on  $J(C_h ; \infty^\pm)$. As
$S^+(\infty^+)=\infty^-$ we see that  $S^+$--real point of
$\phi^{-1}(p)$ is obtained by taking the limit
$Q \mapsto \infty^+$ in $S^+ (Q) + Q$ along an appropriate
real analytic curve on $\breve{C}_h$. At last from the
computation of the action of $S^+$ on $\Lambda $ we get
$S^+ (\Lambda_2) = \Lambda_2$ which shows that the $S^+$--real
part of $(\phi^{-1}(p),S^+)$ is always a circle $\RR / \Lambda_2$.
This gives the last two lines in the table.   {\hfill $\Box $}


\section{The Lagrange top and the non--linear
	 Schr\"odinger equation}

      Our final remark concerns a previously unknown relation
between the real solutions of the Lagrange top and the one--gap
solutions of the nonlinear Schr\"odinger equation
$$
	u_{xx}	=  i u_t \pm 2 |u|^2 u	 \ .
   \leqno{\big(NLS^\pm}\big)
$$
In the physical applications both forms of (NLS) are of interest.
Comparing Theorem \ref{th.Lagrange} to the results of Previato
\cite{Previato} we note that the invariant manifolds of
one--gap solutions of the NLS equation are isomorphic to the
invariant manifolds of the Lagrange top. This relation can be
made explicit if we compare the expressions for the
solutions found in Theorem \ref{theta.th} to the well known
formulae for $u(x,t)$ \cite{Belokolos,Previato}. We shall see
that the $S^\pm$--real solutions of the Lagrange top give also
one--gap solutions of $NLS^\pm$ equation. Recall that, according to the
preceding section, a $S^-$--real solution is an usual real solution of the
Lagrange top (\ref{Ltop}), and that a $S^+$--real solution is a real solution
of the system (\ref{Ltop*}).

 Let $X_E$,
$X_{\Omega_3}$ be the Hamiltonian vector fields (\ref{Ltop})
and (\ref{Ltop1}) respectively and put
$$
      \frac{\partial}{\partial x} = \tfrac{1}{2} X_E ,
      \qquad
      \frac{\partial}{\partial t}
    = \tfrac{1}{4} (m-1) \Omega_3 X_E
      + \tfrac{1}{8} \big( 2 h_3 - (3 m +1)\Omega_3^2 \big)
	X_{\Omega_3} \ .
$$
As $\frac{\partial}{\partial x}$ and $\frac{\partial}{\partial t}$ define
translation invariant vector fields  on the generalized Jacobian
$J(C_h;\infty^\pm)$ then fixing an arbitrary point for
origin we may introduce $(x,t)$ coordinates on
$J(C_h;\infty^\pm)$ (and hence on the complex invariant manifold $T_h$). 
If the real part $T_h^{\RR}$ of $T_h$ is not empty, then we shall chose for
origin a real point. As the real vector fields $\frac{\partial}{\partial x}$
and $\frac{\partial}{\partial t}$ are tangent to the Liouville torus
$T_h^{\RR}$, then $(x,t)$ provide real affine coordinates on it.
Denote at last by
$u^-(x,t)$  the restriction of the function $\overline{\epsi} \Omega_1 +
\epsi \Omega_2$ on  the Liouville torus
$T_h^{\RR}$ of the
Lagrange top (\ref{Ltop}). 

Similarly, let 
$u^+(x,t)$ be the restriction of the function $\overline{\epsi} \Omega_1 +
\epsi \Omega_2$ on a connected component of the $S^+$--real part of 
$J(C_h;\infty^\pm)$. If the origin belongs to this component too, then as
above we conclude that $x, t \in \RR$.

\begin{pr}
  \label{sch.pr}
The functions $u^+(x,t)$ and $u^-(x,t)$ satisfy $NLS^+$ and
$NLS^-$ respectively.
\end{pr}

       The proof of the above Proposition is a straightforward
computation (compare with \cite{Previato}, Theorem 2.2).  From
the definition of $u^\pm$ we get $\overline{u}^- =
\overline{\epsi} \Omega_2 + \epsi \Omega_1$ and
$\overline{u}^+ = - \overline{\epsi} \Omega _2 - \epsi
\Omega _1\,$. It follows that
$|u^\pm|^2 = \mp (\Omega_1^2 + \Omega_2^2)$ and it is easy to
check that
$$
       u_{xx}^\pm = i u_t^\pm  \pm2  |u^\pm|^2 u^\pm
$$
is equivalent to the system
\begin{eqnarray*}
	  (\Omega_1)_{xx} + (\Omega_2)_t
      &=& \pm 2 \Omega_1 (\Omega_1^2 + \Omega_2^2)
  \\	  (\Omega_2)_{xx} - (\Omega_1)_t
      &=& \pm 2 \Omega_2 (\Omega_1^2 + \Omega_2^2)
\end{eqnarray*}
where $\Omega_1 , \Omega_2$ are defined on the $S^\pm$--real
part of $T_h$ respectively. Using (\ref{Ltop}) we get for the
derivatives along $X_E$
$$
	 \bpp{\Omega}_1 + \, (m-1) \Omega_3 \bp{\Omega}_2 \
   \ = \ -m \Omega_1 \Omega^2_3 - \Omega_1 \Gamma_3
$$
and as
$$
      \Gamma_3 \ = \ \tfrac{1}{2}\big( \Omega_1^2 + \Omega_2^2
		    + (1+m) \Omega_3^2 \big) - E \ .
$$
then
$$
       \bpp{\Omega }_1 + \, (m-1) \Omega_3  \bp{\Omega}_2 \ \
  = \ -\tfrac{1}{2} \Omega _1 (\Omega _1^2 + \Omega_2^2)
      + \Omega_1 \big( E - \tfrac{3m+1}{2}\Omega_3^2 \big) \ .
$$
At last as $X_{\Omega_3} \Omega_2 = - \Omega_1$ we conclude that
\begin{eqnarray*}
	    ( \Omega_1)_{xx} + (\Omega_2)_t
     &=&  - 2 \Omega_1 (\Omega_1^2 + \Omega_2^2)
  \\	    ( \Omega_2)_{xx} - (\Omega_1)_t
     &=&  - 2 \Omega_2 (\Omega_1^2 + \Omega_2^2) \; .
\end{eqnarray*}
This proves also that $u^+$ is a solution of $NLS^+$
(we have just to substitute $\Omega_1 \mapsto i\Omega_1$,
$\Omega_2 \mapsto i \Omega_2$).
{\hfill $\Box $}


\newpage

\appendix

\section{Appendix: Linearization of the Lagrange top on
		   an elliptic curve}
  \label{appendix}

     The purpose of the present Appendix is to give a brief
account of some ``well known'' facts concerning the
linearization of the Lagrange top on an elliptic curve. All
algebraic varieties below come equipped with real
structures. We shall make the following convention. If the
complex algebraic varieties $V_1$ and $V_2$ are isomorphic
over $\RR$, then we shall simply write $V_1 = V_2$.

    Further we shall suppose that the  invariant complex
level set
$$
      T_h = \big\{ \,
	      (\Omega,\Gamma ) \in \CC^6 :   \,
	     H_1 = 1\,, \, H_2 = h_2\,, \, H_3 = h_3\,, \,
	     H_4 = h_4	\, \big\}
$$
of the Lagrange top (\ref{Ltop}) is smooth, and moreover
$h = (h_2, h_3, h_4) \in \RR^3$. Thus $T_h$ has a natural
real structure, and if $T_h^\RR$ is its real part we make the
assumption $T_h^\RR \neq \emptyset$. Recall that to $T_h$
we associate the following smooth algebraic curves

    i) the Lagrange curve
       $\Gamma_h = \{\eta^2 = 4 \xi^3 -g_2 \xi - g_3 \}$
       where  $g_2 = g_2(h)$, $g_3 = g_3(h)$ are given by
       (\ref{g2g3}) and (\ref{eah}). The polynomial
       $4 \xi^3 - g_2 \xi - g_3$ has three real roots, so the
       curve $\Gamma_h$ has two ovals. Denote by
       $\overline{\Gamma}_h$ the completed curve $\Gamma_h$.

   ii) the spectral curve
       $\tilde{C}_h = \{\mu^2 + f(\lambda) = 0\}$ of the
       Lax pair of Adler and van Moerbeke (\ref{LaxL}),
       with the natural anti--holomorphic involution
       $(\lambda,\mu) \mapsto
       (\overline{\lambda},\overline{\mu})$,
       where $f(\lambda)$ is given by (\ref{fh}). It is isomorphic
       over $\RR$ to the curve $C_h=\{\mu^2=f(\lambda)\}$ with an
       anti--holomorphic involution $(\lambda,\mu) \mapsto
       (\overline{\lambda},-\overline{\mu})$, so $\tilde{C}_h=C_h$.
       The polynomial $f(\lambda)$ has two pairs of complex
       conjugate roots.

  iii) the Jacobian $J(C_h) = {\rm Pic}^2(C_h)$ of $C_h$ which is
       identified, via the Euler--Weil map (\cite{Weil}),
       to the Lagrange curve $\overline{\Gamma}_h$, so
       $J(C_h) = \overline{\Gamma}_h$.

       According to the context the curves $\tilde{C}_h$, $C_h$
will be considered either as affine, or as completed and
normalized curves.

       Recall also that the generalized Jacobian
$J(C_h;\infty^\pm) = \CC^2/\Lambda$ of the elliptic curve
$C_h$ with two points identified is defined as an extension of
$J(C_h)$ by $C^*$
$$
	0 \stackrel{exp}		 {\rightarrow}
    \CC^* \stackrel{\iota}		 {\rightarrow}
    J(C_h;\infty^\pm)\stackrel{\phi}	 {\rightarrow}
    J(C_h)				  \rightarrow
    0 \ .
$$
By Theorem \ref{th.Lagrange} the invariant complex level set
$T_h$ is identified to $J(C_h ; \infty^\pm) - D_\infty$,
where $D_\infty = \phi^{-1}(p)$, $p=\infty \in \Gamma_h$,
so we obtain the following exact sequence
\begin{equation}
  \label{ext4}
	0  \stackrel{exp}		  {\rightarrow}
     \CC^* \stackrel{\iota}		  {\rightarrow}
      T_h  \stackrel{\phi}		  {\rightarrow}
      \Gamma_h				   \rightarrow
      0 \; .
\end{equation}
Denote by $\overline{T}_h$ the variety $T_h$ completed by
the curve $D_\infty$, so $\overline{T}_h=J(C_h;\infty^\pm)$.
It follows from Theorem \ref{sigma.th} that a point
$t\in J(C_h)$ is defined by $\Gamma_3(t)$ and its
derivative in $t$, and hence
\begin{equation}
  \label{fi}
	\phi : T_h			   \rightarrow
	\Gamma_h :(\Omega,\Gamma)	   \mapsto
	(\eta, \xi) \,, 			\ \
	\xi  = - \tfrac{1}{2} \Gamma_3 \,,	\ \
	\eta = - \tfrac{1}{2} \tfrac{d}{dt} \Gamma_3(t)
	     = - \tfrac{1}{2}(\Gamma_1 \Omega_2
	       - \Gamma_2 \Omega_1) \ .
\end{equation}
The map $\iota$ in (\ref{ext4}) defines a $\CC^*$--action on
$T_h$ which is just the action of the linear complex flow
of (\ref{Ltop1}). The latter is obviously given by
\be
  \label{cstar}
   & &	\Omega_1 \pm i\Omega_2		    \mapsto
	 e^{\pm b}(\Omega_1 \pm i\Omega_2)	  \qquad
	 (M_3, \Gamma_3)		    \mapsto
	 (M_3, \Gamma_3)
\\ \nonumber
   & &	 \Gamma_1 \pm i\Gamma_2 	    \mapsto
	 e^{\pm b}(\Gamma_1 \pm i\Gamma_2)     \  \qquad
	 e^{b} \in \CC^*			  \ .
\ee
This  $C^*$--action is free and compatible with the
projection map $\phi$ so we have a well defined quotient
map
$$
      \phi : \, T_h / \CC^*    \rightarrow   \Gamma_h  \ .
$$
which is an isomorphism. It is obviously prolonged to the
isomorphism
$$
      \phi : \, \overline{T}_h / \CC^*	     \rightarrow
		\overline{\Gamma}_h	     \ .
$$
As $\Gamma_3$ is a first integral of (\ref{Ltop1}), then
the corresponding flow is projected on
$\overline{\Gamma}_h$ to the identity. According to
Theorem \ref{sigma.th} we have
$\Gamma_3(t) = -2 \wp(t) + {\rm constant}$, and hence the flow
of the Lagrange top is projected to a linear flow on the Lagrange
curve $\Gamma_h$. The real part of $T_h$ is a torus
$T_h^\RR \sim S_1 \times S_1$ on which the real flow of
(\ref{Ltop1}) defines a free circle action $\Re = S^1$
compatible with $\phi$. $T_h^\RR$ is compact and connected
so is $\phi(T_h^\RR)$. It follows that
$\phi \big( T_h^\RR \big) = \phi \big( T_h^\RR / \Re \big)$ is
contained in the compact oval of the Lagrange curve $\Gamma_h$.
In fact, $\phi$ provides an isomorphism between $T_h^\RR/ \Re$
and this oval. Indeed, the only thing we need to check is
that the pre--image of a point on this compact oval,
under the map $\phi : T_h^{\RR} \rightarrow \Gamma_h$,
is a single orbit of the system (\ref{Ltop1}), that is to say
a circle. But a point $t$ on $\Gamma_h$ is determined by
$\Gamma_3(t)$ and $\frac{d}{dt} \Gamma_3(t)
 = \Gamma_1 \Omega_2 - \Gamma_2 \Omega_1 $. This combined
with the first integrals amounts to fix $\Omega_3, \Gamma_3$,
the lengths
$$
     \Omega_1^2  +  \Omega_2^2		 \ , \qquad
     \Gamma_1^2  +  \Gamma_2^2		 \ ,
$$
the scalar product
$$
     \Omega_1 \Gamma_1	+  \Omega_2 \Gamma_2
$$
and the vector product
$$
     \Gamma_1 \Omega_2	-  \Gamma_2 \Omega_1
$$
of the real vectors $(\Omega_1,\Omega_2)$,
$(\Gamma_1,\Gamma_2)$, which defines a circle. To summarize,
we have

\begin{th1}{\rm (}Lagrange linearization{\rm )}
\label{th.A}

$ \,$ i{\rm )}	$\phi : T_h/\CC^*	       \rightarrow
	       \Gamma_h$ is an isomorphism

 $ $ ii{\rm )} $\phi : \overline{T}_h/\CC^*   \rightarrow
		\overline{\Gamma}_h$ is an isomorphism

    iii{\rm )} the image of the flow of {\rm (\ref{Ltop1})} on
	       $\overline{\Gamma}_h$ is the identity, and the
	       one of {\rm (\ref{Ltop})} is linear.

 $  $ iv{\rm )}  the map $\phi$ provides an isomorphism between
	       $T_h^{\RR}/\Re$ and the compact oval
	       of the affine real curve $\Gamma_h$.
\end{th1}

     The above theorem may be attributed to Lagrange
\cite[p.254]{Lagrange} who computed the differential equation
satisfied by the nutation $\Gamma_3(t)$. It worth noting
that this computation was published in 1813 (the year when
Lagrange died) by Poisson \cite{Poisson} as completely new,
and without mentioning Lagrange.

     There is another more sophisticated way to linearize the
Lagrange top on the elliptic curve $\Gamma_h$, by making use
of the Lax pair representation (\ref{LaxL}) \big(see
\cite{Adler,Ratiu,Verdier,Audin1,Audin2}\big)
\ben
       \frac{d}{dt} \,
	 \big( \lambda^2 \chi + \lambda M -\Gamma \big)
     = \big[ \, \lambda^2 \chi+\lambda M -\Gamma \,,\,
	 \lambda \chi+ \Omega \, \big] \ .
\een
Namely, let $\stackrel{\circ}{C}_h$ be the affine curve
$\tilde{C}_h$ with its Weierstrass points removed (they
correspond to the roots of $f(\lambda)$\big), and put
$A(\lambda) = \lambda^2 \chi+\lambda M -\Gamma$. As
$-\mu\big(\mu^2 + f(\lambda)\big) = {\rm det}\,
(A(\lambda)-\mu I)$,
then for $(\lambda,\mu)\in \, \stackrel{\circ}{C}_h$ we
have ${\rm dim \, Ker}\, \big( A(\lambda)-\mu I \big) = 1$.
It follows that the variety
$$
    \Big\{ \,
      (\lambda,\mu)\in \, \stackrel{\circ}{C},	 \,
      [v_0,v_1,v_2]\in \CC \PP^2 :		 \,
      (v_0,v_1,v_2) \in
      {\rm Ker} \, \big( A(\lambda)-\mu I \big)  \,
    \Big\} \,
  \subset \,
    \tilde{C}_h \times \CC \PP^2
$$
is smooth and it is easy to check that its closure in
$\{ \tilde{C}_h \cup \infty^+ \cup \infty^-\} \times \CC \PP^2$
is also  smooth, so we have a holomorphic line bundle on the
compactified and normalized  curve
$\{ \tilde{C}_h \cup \infty^+ \cup \infty^-\}$ (this also follows from 
\cite[Proposition 2.2]{gv}). One computes further that the degree of this
bundle is four and there is always a meromorphic section with a pole divisor
$D = R_+ + R_- + \infty^+ + \infty^-$. Of course, the divisor
$D$ depends on the coefficients of the polynomial matrix
$A(\lambda)$, and hence on $(\Omega,\Gamma)$. Consider now
the map
\ben
    \tilde{\phi} \ : \quad T_h	\ \ &  \rightarrow  &
       {\rm Pic}^2(\tilde{C}_h)
    =  J(\tilde{C}_h)
    =  \overline{\Gamma}_h
 \\
       (\Omega,\Gamma)		    &	 \mapsto       &
       [R_++R_-]
\een
where the divisor
$R_\pm = \big( \lambda(R_\pm),\mu(R_\pm) \big) \in \tilde{C}_h$
equals to
$$
       \lambda(R_\pm)
     = \frac{\Gamma_1 \mp i \Gamma_2}
	    {\Omega_1 \mp i\Omega_2} \,,	  \quad
	\mu(R_\pm) =  \pm i \, \big(-\Gamma_3
		    + (1+m)h_4 \lambda(R_\pm)
		    + \lambda^2(R_\pm) \big) \ .
$$
Note that, according to Theorem \ref{sigma.th}, the map
$\tilde{\phi}$ is prolonged to a holomorphic map
$$
	\tilde{\phi} : \, \overline{T}_h      \rightarrow
	{\rm Pic}^2(\tilde{C}_h) = J(\tilde{C}_h)
			   = \overline{\Gamma}_h \ .
$$
We shall show that the map $\tilde{\phi}$ provides a
linearization of the Lagrange top on $\overline{\Gamma}_h$. It
is obvious that $\tilde{\phi}$ is compatible with the
$\CC^*$ action (\ref{cstar}) on $T_h, \tilde{T}_h$, so we
have the holomorphic maps
$$
      \tilde{\phi} : \, \overline{T}_h/\CC^*  \rightarrow
			\overline{\Gamma}_h   \,,   \qquad
	      \phi : \, \overline{T}_h/\CC^*  \rightarrow
			\overline{\Gamma}_h   \,,   \qquad
	 \tilde{\phi} \circ \phi^{-1} : \,
			\overline{\Gamma}_h   \rightarrow
			\overline{\Gamma}_h   \ .
$$
Remembering that $\overline{\Gamma}_h $ is a complex torus, we
conclude that if $z \in \CC/\Lambda \sim \overline{\Gamma}_h$,
then $\tilde{\phi}\circ \phi^{-1}(z)= kz$, for some
$k \in \ZZ$, and hence $\tilde{\phi}$ provides a linearization
on $\overline{\Gamma}_h$ too.
The map $\tilde{\phi}$ is a non--ramified covering of degree $k^2$
and it is easy to check that $k^2=4$. Indeed, if $R_++R_-$
is linearly equivalent on $\tilde{C}_h$ to $ \infty^++\infty^-$,
then $R_+ = \sigma(R_-)$, where
$\sigma(\lambda , \mu) = (\lambda, -\mu)$ is the elliptic
involution. It follows that
$$
      \frac{\Gamma_1 + i \Gamma_2}{\Omega_1 + i\Omega_2}
    = \frac{\Gamma_1 - i \Gamma_2}{\Omega_1 - i\Omega_2} \
      \Longleftrightarrow \
      \Omega_1 \Gamma_2 - \Omega_2 \Gamma_1
    = \frac{d}{dt} \Gamma_3(t) = 0 \; ,
$$
which shows that the pre--image of the	divisor class
$\infty^+ + \infty^-$  on  $\overline{\Gamma}_h$ with respect
to $\tilde{\phi} \circ \phi^{-1}$ are the four Weierstrass
points on $\overline{\Gamma}_h$. At last we note that
$\tilde{\phi}(T_h^{\RR}/\Re)$, as before, is contained in an
oval of $\overline{\Gamma}_h$. In this case, however,
$\tilde{\phi}$ provides a double non--ramified covering
of $T_h^{\RR}/\Re$ to its image -- the	oval of the curve
$\overline{\Gamma}_h = {\rm Pic}^2(\tilde{C}_h)$ containing
the point $\infty$. Indeed, note that the divisor class of
$\infty^++\infty^-$ represents a real point on
${\rm Pic}^2(\tilde{C}_h)$. It has exactly two real pre--images:
the two Weirstrass points contained in the compact oval of
$\Gamma_h$, and the remaining two Weirstrass points
are not real. Thus we proved the following

\begin{th1}  \label{th.B}
      {\rm (}Linearization by making use of a Lax pair{\rm )}
$\,$ Let $\Gamma_h$ be the affine curve defined above, and
\vmmm
\ben
	  \stackrel{\circ}{T}_h
   \ = \  T_h \setminus \big\{(\omega,\Gamma)\in \CC^6:
	 \Omega_1 \Gamma_2 - \Omega_2 \Gamma_1
     =	0   \big\}    \ .
\een
\vspace{-15mm} $$ $$
Then

 $\,$  i{\rm )}
       $\tilde{\phi} : \, \stackrel{\circ}{T}_h/\CC^*
       \rightarrow
       \Gamma_h$ is a non--ramified covering of degree four

  \  ii{\rm )}
       $\tilde{\phi} : \, \overline{T}_h/\CC^*
       \rightarrow
       \overline{\Gamma}_h$ is a non--ramified covering
       of degree four

 iii{\rm )}
      the image of the flow of {\rm (\ref{Ltop1})} on
      $\overline{\Gamma}_h$ is the identity, and the one of
      {\rm (\ref{Ltop})} is linear

\, iv{\rm )}
     the map $\tilde{\phi}$ provides a double non--ramified
     covering of $T_h^{\RR}/\Re$ to its image -- the  oval
     of the compactified and normalized curve
     $ \overline{\Gamma}_h = \Gamma_h  \cup \infty$
     containing the point $\infty$.
\end{th1}

Statement iv) is due to M.Audin. In this form it appeared
first in \cite{Audin1} (Proposition 3.3.2) but the proof
is not correct. Earlier Verdier \cite{Verdier} wrongly claimed that
the map $\tilde{\phi}$ provides an isomorphism between
$T_h^{\RR}/\Re$ and its image. Statement iii) is a well known
fact which, however, seems to be never rigorously proved.
Thus Adler and van Moerbeke \cite{Adler} and then Ratiu
and van Moerbeke \cite{Ratiu} proposed a  ``proof"  based on
a general scheme for linearizing the flow defined by a Lax
pair with a spectral parameter (e.g. Adler and van
Moerbeke \cite{Adler}, Theorem 1, p.337). The Lax pair
(\ref{LaxL}) does not fit, however, to the general
procedure, as its spectral curve is always reducible.
Of course this is only a minor technical difficulty
as we may also use the Lax pair (\ref{Lax1}). It was
proposed in \cite[p.351]{Adler} and \cite{Ratiu} to
consider instead of  the Lax pair (\ref{LaxL}), another Lax
pair
\vmm
\begin{equation}
  \label{lam}
      \frac{dA^\epsilon}{dt} = [A^\epsilon,B^\epsilon]
\end{equation}
\vspace{-13mm} $$ $$
where in the notations of \cite{Adler} we have
$$
     A^\epsilon  =  A^\epsilon(h)  =
	\left(
	   \begin{array}{ccc}
	      \epsi h^2  &  \beta   &  i \beta^*  \\
	      - \beta^*     & -\omega  &  0	     \\
	      i \beta	    &	 0     &  \omega
	   \end{array}
       \right) \ ,			   \qquad
      B^\epsilon = B^\epsilon(h)
		 = \frac{1}{I_1} \, [h^{-1}A^\epsilon(h)]_+
	       \quad ,
$$
$[\,.\,]_+$  means ``polynomial part'' and
\begin{align}
    \beta & =  y + hx					       &
	y & = \tfrac{1}{\sqrt{2}} \,  (\gamma_1 - i \gamma_2)  &
	x & = \tfrac{I_1}{\sqrt{2}}\, (\Omega_1 - i \Omega_2)  &
\nonumber \\
   \beta ^*   & = \overline{y}+ h\overline{x}		       &
 \overline{y} & = \tfrac{1}{\sqrt{2}} \,(\gamma_1 + i \gamma_2)&
 \overline{x} & = \tfrac{I_1}{\sqrt{2}}\,(\Omega_1 + i\Omega_2)&
 \nonumber \\
	      & 					       &
     i \omega & =  z_0 I_1 h^2 + I_3 \Omega_3 h + \gamma_3  \ .&
	      & 					       &
\nonumber
\end{align}
To obtain from the notations of \cite{Adler} our notations we
just replace
\vm
\ben
       \gamma_i   =  -\Gamma_i \ ,	  \quad
       z_0 = I_1  =  1	       \ ,	  \quad
       I_3 = 1+m	       \ ,	  \quad
	h  = \lambda	       \ .
\een
\vspace{-12mm} $$ $$
For the spectral curve $X_\epsilon$ of $A^\epsilon$ we obtain
\vmm
\be
       {\rm det} \, \big( A^\epsilon(h) - zI \big)
     &=& (\epsilon h^2-z)(z^2-\omega^2) - 2 \beta \beta^* z
\nonumber \\
\label{xep}
     &=&  - z^3
	  + \epsilon h^2 z^2
	  + (-2 \beta \beta^*+ \omega^2)z
	  - \epsilon h^2 \omega^2
\\ \nonumber
     &=&  0 \ .
\ee
\vspace{-15mm} $$ $$
This is generically a smooth irreducible genus four curve, so
the Lax pair (\ref{lam}) fits to Theorem~1, p.337 in
\cite{Adler}. Thus the flow of (\ref{lam}) linearizes on
${\rm Jac}\,(X_\epsilon)$ and when $\epsilon \rightarrow 0$
it goes over into a linear flow on the compact piece of
${\rm Jac} \, (X_0)$ which is just the Lagrange elliptic curve.
On the other hand  the differential equation (\ref{lam}) for
$\epsilon=0$ is, modulo a linear change of the variables, the
original system (\ref{Ltop}) which establishes once again
Theorem \ref{th.B}, ii). It is easy to see, however, that the
above approach does not work as for $\epsilon \neq 0$
the Lax pair (\ref{lam}) {\it does not define a
differential equation}. Indeed, note that (\ref{lam})
is equivalent to the Lax pair
\vm
\begin{equation}
  \label{lam1}
     \frac{dA^0}{dt}  = [A^0,B^0] - \frac{\epsilon h}{I_1}
     \left(
	   \begin{array}{ccc}
	     0	      &      y	    &	  i \overline{y}   \\
	-\overline{y} & i \gamma_3  &		0	   \\
	    i y       &      0	    &	 -i \gamma_3
	    \end{array}
      \right)
\end{equation}
\vspace{-11mm} $$ $$
Its $(1,2)$ entry is computed to be
\vmmm
\ben
      \frac{d\beta}{dt}
    = \frac{i}{I_1}\, \big( y I_3 \Omega_3
		       - x \gamma_3 + h z_0 I_1 y \big)
      - \frac{\epsi h\,y}{I_1}
\een
\vspace{-13mm} $$ $$
and the $(3,1)$ entry is
\vmmm
\ben
	i \, \frac{d\beta}{dt}
      = \frac{1}{I_1}\, \big( - y I_3 \Omega_3
		       + x \gamma_3 - h z_0 I_1 y \big)
	+ \frac{\epsilon^3 \, h \, y}{I_1}
\een
\vspace{-13mm} $$ $$
so $y \equiv 0$ and in a similar way $\overline{y} \equiv 0$.

      More generally, it is seen from the coefficients of the
spectral curve $X_\epsilon \,$, $\epsilon \neq 0$, that the
functions
$$
     \Omega_1^2 + \Omega_2^2		     \qquad \
     \gamma_1^2 + \gamma_2^2		     \qquad \
     \Omega_1 \gamma_1 + \Omega_2\gamma_2    \qquad \
     \gamma_3				     \qquad \
     \Omega_3
$$
are invariants for {\it any } isospectral deformation of the
matrix $A^\epsilon$. By continuity these five functions are
invariants for $\epsilon = 0$ too, so the vector field in
$\CC^6$ obtained as $\epsilon \rightarrow 0$ is collinear
to the linear vector field of (\ref{Ltop1}). Of course
there is no analytic change of variables in $\CC^6$ which
sends the orbits of (\ref{Ltop1}) to orbits of (\ref{Ltop}).

\end{document}